\documentclass[a4paper,11pt]{article}
\pdfoutput=1 
\usepackage{jheppub}
\usepackage{rotating}
\usepackage{multicol}
\usepackage{graphicx}
\usepackage{booktabs}
\usepackage{amssymb,bm,mathrsfs,bbm,amscd}
\usepackage{subfig,color,slashed}
\usepackage{amssymb}
\usepackage[normalem]{ulem}
\usepackage[utf8]{inputenc}
\usepackage{ulem}
\usepackage{lscape}
\usepackage{lipsum}

\def\lsim{\mathrel{\rlap{\lower4pt\hbox{\hskip1pt$\sim$}}
    \raise1pt\hbox{$<$}}}         
\def\gsim{\mathrel{\rlap{\lower4pt\hbox{\hskip1pt$\sim$}}
    \raise1pt\hbox{$>$}}}         

\title{Light gauge boson interpretation for $(g-2)_\mu$ and the $K_L \rightarrow \pi^0 + \text{(invisible)}$ anomaly at the J-PARC KOTO experiment}

\author[\dagger \star]{Yongsoo Jho,}
\author[\dagger]{Sung Mook Lee,}
\author[\dagger]{Seong Chan Park,}
\author[\dagger]{Yeji Park}
\author[\dagger]{and Po-Yan Tseng}

\affiliation[\dagger]{Department of Physics and IPAP, Yonsei University, \\
Seoul 03722, Republic of Korea}

\affiliation[\star]{Center for Theoretical Physics of the Universe, Institute for Basic Science, \\
Daejeon 34126, Republic of Korea}

\emailAdd{jys34@yonsei.ac.kr}
\emailAdd{sungmook.lee@yonsei.ac.kr}
\emailAdd{sc.park@yonsei.ac.kr}
\emailAdd{yeji.park@yonsei.ac.kr}
\emailAdd{tpoyan1209@gmail.com}

\abstract{
We discuss a list of possible light gauge boson interpretations for the long-standing experimental anomaly in $(g-2)_\mu$ and also recent anomalous excess in $K_L \rightarrow \pi^0 + \text{(invisible)}$ events at the J-PARC KOTO experiment. We consider two models: \emph{i)} $L_\mu - L_\tau$ gauge boson with heavy vector-like quarks and \emph{ii)} $(L_\mu - L_\tau) + \epsilon (B_3 - L_\tau)$ gauge boson in the presence of right-handed neutrinos. When the light gauge boson has mass close to the neutral pion in order to satisfy the Grossman-Nir bound,  the models successfully explain the anomalies simultaneously while satisfying all known experimental constraints. We extensively provide the future prospect of suggested models.
}

\begin{document} 
\maketitle
\flushbottom

\section{Introduction}

\ \ \ \ \ The KOTO experiment at the Japan Proton Accelerator Research Complex (J-PARC) recently released their result on $K_L \to \pi^0 \nu \bar{\nu}$ searches~\cite{Shinohara:2019, Tung:2019, Lin:2019}: four candidate events were observed in the signal region over the background estimation $0.05\pm 0.02$. One of those candidate events is still suspected as a background from overlapped pulse, but other three events are distinctive in their properties from the known backgrounds.

The required branching ratio of $K_L \to \pi^0 \nu \bar{\nu}$ for the three candidate events is~\cite{Kitahara:2019lws}
\begin{eqnarray}
{\rm Br}(K_L \to \pi^0 \nu \bar{\nu})_{\rm KOTO}=2.1^{+2.0}_{-1.1}
\times 10^{-9}\,,
\end{eqnarray}
if we assume these events come from $\pi^0 \nu \bar{\nu}$ decay channel. On the other hand, the Standard Model (SM) prediction for this channel mainly from the penguin and box diagrams is~\cite{Buras:2006gb,Brod:2010hi,Buras:2015qea}
\begin{eqnarray}
{\rm Br}(K_L \to \pi^0 \nu \bar{\nu})_{\rm SM}= (3.00 \pm 0.30) \times 10^{-11}\,,
\end{eqnarray}
and it is about two orders of magnitude smaller than the KOTO events requirement.

At the same time, the NA62 updated their result on $K^+ \to \pi^+ \nu \bar{\nu}$ and provided the 95\% CL upper limit~\cite{Ruggiero:2019}
\begin{eqnarray}
{\rm Br}(K^+ \to \pi^+ \nu \bar{\nu})_{\rm NA62}<2.44
\times 10^{-10}\,,
\end{eqnarray} 
which is consistent with SM prediction ${\rm Br}(K^+ \to \pi^+ \nu \bar{\nu})_{\rm SM}= (9.11 \pm 0.72) \times 10^{-11}$~\cite{Buras:2006gb,Buras:2015qea}. The $K_L$ and $K^+$ decay branching ratios are strongly connected through the Grossman-Nir (GN) bound~\cite{Grossman:1997sk}, which requires
\begin{eqnarray}
{\rm Br}(K_L \to \pi^0 \nu \bar{\nu})\leq 4.3 \times {\rm Br}(K^+ \to \pi^+ \nu \bar{\nu})\,.
\end{eqnarray}
However, there is a method to circumvent the GN bound~\cite{Fuyuto:2014cya,Kitahara:2019lws}, because of detail experimental arrangement 
and large background from $K^+ \to \pi^+ \pi^0$ for 
$K^+ \to \pi^+ \nu \bar{\nu}$ measurement. 
As a result, at NA62 ${\rm Br}(K^+ \to \pi^+ \nu \bar{\nu})$ measurement, 
the kinematic region of the missing mass $100 \text{ MeV} < m_{\rm miss} <165$ MeV around $\pi^0$ mass was overlooked. 
Therefore, if a resonance particle is carrying mass around this window, 
which produced from Kaon decay 
and then decays into $\nu \bar{\nu}$ 
(or invisibly decays into other dark sector particles), 
the GN bound would be significantly weakened.
Consequently, a particle, which couples to both neutrinos and quarks 
with mass around $\pi^0$ mass, 
might provides consistent explanation for both KOTO and NA62 results. Recently, intriguing explanations of KOTO event excess with the models including light scalars coupled to quarks \cite{Egana-Ugrinovic:2019wzj, Dev:2019hho, Liu:2020qgx}, light dark sector fermions \cite{Fabbrichesi:2019bmo}, and generic higher dimensional operators in the neutrino sector \cite{Li:2019fhz} have been suggested.

The long-lasting $(g-2)_\mu$ discrepancy at the level of $(3.3 - 4.1)\sigma$ between observations \cite{Brown:2001mga,Bennett:2002jb,Bennett:2004pv,Bennett:2006fi} and SM predictions \cite{Davier:2010nc,Jegerlehner:2011ti,Hagiwara:2011af,Jegerlehner:2017lbd,Davier:2017zfy,Keshavarzi:2018mgv,Davier:2019can,Keshavarzi:2019abf} strongly implies the presence of new physics
\footnote{Caveat: There are ambiguities on theoretical calculations for $(g-2)_\mu$.
For example, the recent lattice method for the hadronic vacuum polarization from Ref.\cite{Borsanyi:2020mff}, their result eliminates the need to invoke new physics to explain the discrepancy between SM prediction  and experimental measurement.
In our work, we have considered the $3.8 \sigma$ muon g-2 discrepancy, 
based on recently updated results by KNT2019~\cite{Keshavarzi:2019abf}.}. 
Various new physics explaining $(g-2)_\mu$ has been suggested so far, and the $U(1)_{L_\mu-L_\tau}$ gauge boson $X$  \cite{Foot:1990mn,He:1990pn,He:1991qd,Baek:2001kca} with its mass in the range $10 \text{ MeV} \lsim m_X \lsim 200$ MeV is still preferred after 
taking the present experimental observations \cite{Bauer:2018onh}. 
The discovery potentials of $X$ gauge boson in many current and future experiments, such as mono-photon ($e^+ e^- \rightarrow \gamma X$) and di-muon ($e^+ e^- \rightarrow \mu^+ \mu^- X$) searches at Belle II \cite{Kaneta:2016uyt,Araki:2017wyg,Jho:2019cxq}, leptonic decay of charged kaons ($K^+ \rightarrow \mu^+ \nu_\mu X$) decays at NA62 \cite{Krnjaic:2019rsv}, future neutrino-trident upper bound ($\nu_{\mu,\tau} N \rightarrow \nu_{\mu,\tau} N \mu^+ \mu^-$) from DUNE \cite{Ballett:2019xoj} and future muon beam experiment \cite{Kahn:2018cqs} have been extensively explored. 
Since the predominating decay channel for $m_X$ 
below muon threshold is neutrino pairs ($\nu_\mu \bar{\nu}_\mu$ and $\nu_\tau \bar{\nu}_\tau$), it becomes plausible explanation for KOTO events, but depends on its couplings to quarks.

In this paper, we check simultaneous explanations of $(g-2)_\mu$ and KOTO events with a light gauge boson $X$. We found that only through the mixing with photon, it cannot generates sufficient ${\rm Br}(K_L \to \pi^0 X)$ for KOTO events excess, meanwhile satisfies other experimental constraint, especially from BaBar \cite{Lees:2017lec} and NA64 \cite{Banerjee:2017hhz} (Section \ref{subsec:Kinetic_mixing}). 
Alternatively, we investigate the two kinds of plausible interactions between a new light gauge boson $X$ coupled to muons and quarks through the followings:

\begin{itemize}
\item {\bf $L_\mu-L_\tau$ gauge boson $X$ with heavy vector-like quarks (VLQs) (Section \ref{subsec:Model1_LmuLtau_w_VLQs})}:

Introducing heavy VLQs at TeV scale couple to both $L_\mu-L_\tau$ gauge boson and SM quark sector \cite{ Altmannshofer:2014cfa} is a promising way to enhance ${\rm Br}(K_L \to \pi^0 X)$. The flavour changing neutral current (FCNC) is generated at tree level due to the VLQs' non-trivial contributions to the off diagonal elements of the quark mass matrix. Then we check the consistency with existing constraints such as Br($K^+ \rightarrow \pi^+ X$), $K^0-\bar{K}^0$ mixing, Br($K_L \to \mu^+ \mu^-$), Cabibbo-Kobayashi-Maskawa (CKM) unitarity.

\item {\bf $(L_\mu - L_\tau) + \epsilon(B_3-L_\tau)$ gauge bosons $X$ in the presence of right-handed neutrinos (RH$\nu$) (Section \ref{subsec:Model2_LmuLtau_B3Ltau_w_RHnu})}: 

We also consider $(L_\mu - L_\tau) + \epsilon(B_3 - L_\tau)$ gauge boson to explain KOTO events ($K_L^0 \rightarrow \pi^0 X$) and check whether the preferred parameter region satisfies existing constraints: Br($K^+ \rightarrow \pi^+ X$), Br($B^+ \rightarrow K^+ X$), $B^0-\bar{B}^0$/$D^0-\bar{D}^0$/$K^0-\bar{K}^0$ mixing, Br($B_{d,s} \rightarrow \mu^+ \mu^-$), Br($K_L \rightarrow \mu^+ \mu^-$). The generic kinetic and mass mixings between two gauge bosons (from $L_\mu - L_\tau$ and $B_3 - L_\tau$ for instance) naturally induce this type of gauge coupling.

\end{itemize}

This paper is organized as following. We write down the decay widths and construct the effective operators in next Section~\ref{sec:decay}. In Section~\ref{subsec:Model1_LmuLtau_w_VLQs}, we describe the $L_\mu -L_\tau$ (with heavy VLQs) model framework, formalism, and plausible constraints. The $(L_\mu - L_\tau) + \epsilon (B_3 - L_\tau)$ model detail and relevant constraints are described in Section~\ref{subsec:Model2_LmuLtau_B3Ltau_w_RHnu}. In the last Section~\ref{sec:summary_conclusion}, we summarized our results.

\section{Decay of $K$ and $B$ mesons with FCNCs}
\label{sec:decay}

\subsection{Decay widths and experimental limits}
\label{subsec:Width_exp_limits}

We focus on the effective FCNC couplings of $X$ boson
\begin{eqnarray}
\mathcal{L}_{\rm eff}\supset 
- g^{\rm eff}_{dsX}[\bar{d}_L\gamma^\mu s_L]X_\mu 
- g^{\rm eff}_{sbX}[\bar{s}_L\gamma^\mu b_L]X_\mu
+\text{h.c.}\,,
\end{eqnarray}
where first term is relevant to the KOTO process, 
 and both terms are correlated to each other under these two model frameworks 
that will be discussed in this work.
These FCNC couplings lead to the branching ratios of rare $K$ and $B$ meson decays as follows \cite{Fuyuto:2014cya, Fuyuto:2015gmk}:
\begin{eqnarray}
{\rm Br}(K^+ \rightarrow \pi^+ X) & = & \frac{m_{K^+}^3}{\Gamma_{K^+}} \frac{|g_{dsX}^{\rm eff}|^2}{64\pi m_X^2} \Bigl [ \lambda_1 \Bigl (1, \frac{m_{\pi^+}^2}{m_{K^+}^2}, \frac{m_X^2}{m_{K^+}^2} \Bigr ) \Bigr ]^{3/2} \Bigl [ f_+^{K^+ \pi^+} (m_X^2) \Bigr ]^2, \\
{\rm Br}(K_L \rightarrow \pi^0 X) & = & \frac{m_{K_L}^3}{\Gamma_{K_L}} \frac{(\text{Im }g_{dsX}^{\rm eff})^2}{64\pi m_X^2} \Bigl [ \lambda_1 \Bigl (1, \frac{m_{\pi^0}^2}{m_{K_L}^2}, \frac{m_X^2}{m_{K_L}^2} \Bigr ) \Bigr ]^{3/2} \Bigl [ f_+^{K_L \pi^0} (m_X^2) \Bigr ]^2,
\end{eqnarray}
\begin{eqnarray}
{\rm Br}(B^+ \rightarrow K^+ X) & \simeq & \frac{m_{B^+}^3}{\Gamma_{B^+}} \frac{|g_{sbX}^{\rm eff}|^2}{64\pi m_X^2} \Bigl [ \lambda_1 \Bigl (1, \frac{m_{K^+}^2}{m_{B^+}^2}, \frac{m_X^2}{m_{B^+}^2} \Bigr ) \Bigr ]^{3/2} \left ( \frac{0.33}{1 - m_X^2/(38 \text{ GeV}^2)} \right )^2 \,. \ \ \ \ \ \ \ \ 
\end{eqnarray}
where $\lambda_1(x,y,z)\equiv x^2+y^2+z^2-2xy-2xz-2yz$, and
for $K^+$ and $K_L$ mesons the corresponding form factors $f_+^{K^+ \pi^+} (q^2)$, $f_+^{K_L \pi^0} (q^2)$ are close to the unity. 
The branching ratios of $K^+$ and $K_L$ are correlated to each other, 
since the $K_L$ corresponds to only the imaginary part of $g^{\rm eff}_{dsX}$, 
meanwhile $K^+$ is proportional $|g^{\rm eff}_{dsX}|^2$. 
The total widths for these mesons
$\Gamma_{K^+}  =  5.315 \times 10^{-17} \text{ GeV}$,
$\Gamma_{K_L}  =  1.286 \times 10^{-17} \text{ GeV}$,
$\Gamma_{B^+}  =  4.017 \times 10^{-13} \text{ GeV}$ are used to obtain the branching ratios.~\cite{Tanabashi:2018oca}.

For $m_X$ below the muon threshold and no coupling with electron current, 
only neutrino pair decay mode is kinematic allowed. Furthermore the $X$ boson can also decay invisibly into pair of hidden sector light particles. 
And thus in the rest of this paper, 
we assume that the invisible decay mode dominates 
the light $X$ boson decay.
The required effective coupling strength to explain KOTO events excess from above estimation is
\begin{eqnarray}
\label{eq:geff_KOTO}
|\text{Im }g_{dsX}^{\rm eff}|\simeq 1.16 \times 10^{-12}\,. \ \ \ \ \ \ \text{(KOTO desired FCNC coupling for $q^2 = m_{\pi^0}^2$)}\,,
\end{eqnarray}
We set $m_X\simeq m_\pi^0$ to evade the stringent constraint from $\text{Br} (K^+ \to \pi^+ + \text{invisible})$ decay, which is suffered from overwhelming $K^+ \to \pi^+ \pi^0$ background. Therefore, it can satisfy other upper bounds from current observations of rare $K$ and $B$ meson FCNC decays. 
Taking $K^+$ for example, because of the huge $K^+ \to \pi^+\pi^0$ background, when the square of missing energy around pion mass $q^2\simeq m^2_\pi$, weaker bound 
\begin{eqnarray}
|g_{dsX}^{\rm eff}|\leq 1.256\times 10^{-11} \ \ \ \ \ \  \text{(From Br($K^+ \rightarrow \pi^+ X$) upper limit for $q^2 = m_{\pi^0}^2$)}
\end{eqnarray} 
comes from
$\text{Br} (K^+ \rightarrow \pi^+ + \text{invisible})
|_{q^2 \simeq m_{\pi^0}^2}  \leq  5.6 \times 10^{-8}$ of E949 at BNL \cite{Artamonov:2009sz}.\footnote{Similarly, the recent NA62 result does not provide a significant upper bound in this region ($0.015\,{\rm GeV^2}\lesssim q^2 \simeq m_{\pi^0}^2\lesssim 0.0225\,{\rm GeV^2}$), due to the huge background $K^+ \rightarrow \pi^+ \pi^0$ \cite{Ruggiero:2019}.}
And thus weaker GN bound, i.e. 
$\text{Br} (K_L \rightarrow \pi^0 + \text{invisible})|_{q^2 \simeq m_{\pi^0}^2}\leq 4.3\, \text{Br} (K^+ \rightarrow \pi^+ + \text{invisible})|_{q^2 \simeq m_{\pi^0}^2}$,
can be translated into the limit as
\begin{eqnarray}
|\text{Im} \ g_{dsX}^{\rm eff}|\leq 1.246\times 10^{-11} \ \ \ \ \  \text{(From GN bound for $q^2 = m_{\pi^0}^2$)}
\end{eqnarray}
which is still an order of magnitude larger than the prefer coupling for KOTO events.
The bound for $b$ to $s$ coupling is from
the $\text{Br} (B^+ \rightarrow K^+ + \text{invisible})  \leq  1.3 \times 10^{-5}$ of Belle \cite{Chen:2007zk} and BaBar \cite{delAmoSanchez:2010bk, Lees:2013kla} requires 
\begin{eqnarray}
|g_{sbX}^{\rm eff}| & \leq & 1.23\times 10^{-9} \ \ \ \ \ \ \text{(From Br($B^+ \rightarrow K^+ X$) upper limit for $q^2 = m_{\pi^0}^2$)}\,.
\end{eqnarray}
It may provides additional constraint, 
if the couplings $g^{\rm eff}_{sbX}$ and $g^{\rm eff}_{dsX}$ are correlated.

Explaining the KOTO event excess through $K_L \to \pi^0+X$ with 
light gauge boson $m_X\simeq m_\pi^0$ is still in accordance
with other present experimental constraints.
Further more, if this $X$ boson carries the muonic force 
with coupling strength of $\mathcal{O}(10^{-3})$, 
it can also explain the $(g-2)_\mu$ anomaly~\cite{Jho:2019cxq}.

\subsection{The $U(1)_X$ gauge boson mixing with photon}
\label{subsec:Kinetic_mixing}

\begin{figure}[h]
\centering
\includegraphics[height=3.5in,angle=0]{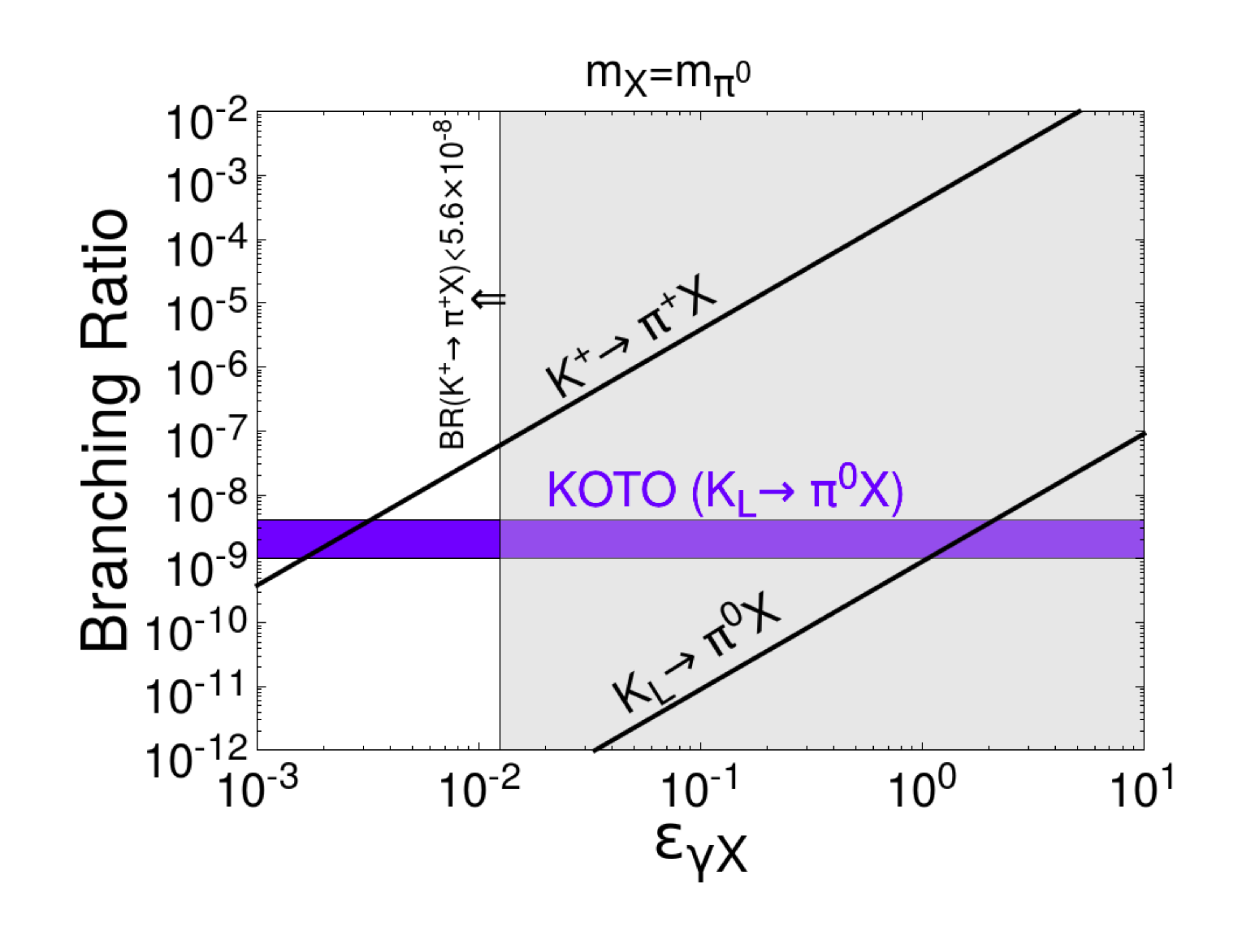}
\caption{\small \label{fig:KL_piX}
The ${\rm Br}(K_L \to \pi^0 + X)$ and ${\rm Br}(K^+ \to \pi^+ + X)$
from kinematic mixing $\epsilon_{\gamma X}$ between $X$ boson and photon. The purple band shows the $1\sigma$ region for KOTO event excess.
The gray region is excluded by E949 with ${\rm Br}(K^+ \to \pi^+ + X)\geq 5.6\times 10^{-8}$.}
\end{figure}

One simple and vastly discussed model in the literature is the $U(1)_X$ gauge boson $X$ kinematic mixing with SM photon or $Z$ boson through the mixing parameters $\epsilon_{\gamma X}$ and $\epsilon_{Z X}$, respectively. However, we would like to show that this single model cannot explain the KOTO event excess 
under the constraint from ${\rm Br}(K^+ \to \pi^+ + \text{invisible})$ of E949.

The $X$ boson couples to the SM quark current through the mixing,
and then the FCNC are generated from one-loop $W$ boson and top quark penguin diagram. The down-type FCNC transitions $b \rightarrow s X$ and $s \rightarrow d X$ are given by
\begin{eqnarray}
\mathcal{L}_{\rm eff} & \supset & \sum_{i=u,c,t} V_{ib} V_{is}^* \frac{G_F}{\sqrt{2}} \frac{e}{8\pi^2} H_0 (x_i) 
[\bar{s} (q^2 \gamma^\mu - q^\mu \slashed{q}) (1-\gamma_5) b] X_\mu \nonumber \\
& & + \sum_{i=u, c,t} V_{is} V_{id}^* \frac{G_F}{\sqrt{2}} \frac{e}{8\pi^2} H_0 (x_i) 
[\bar{d} (q^2 \gamma^\mu - q^\mu \slashed{q}) (1-\gamma_5) s] X_\mu
\end{eqnarray}
in which $V_{ij}$ is the CKM matrix element, $x_i = \frac{m_i^2}{m_W^2}$ $(i=u,c,t)$. The $q$ is outgoing momentum carried by $X$ gauge boson,
therefore the above vertices are suppressed by ratio $m^2_X/m^2_W$, 
where $m^2_W$ comes from Fermi constant $G_F$. The vertex function $H_0 (x)$ consisting of photon component function $D_0 (x_i)$ and $Z$ component function $\tilde{D}_0 (x_i)$, are characterized by $\epsilon_{\gamma X}$ and $\epsilon_{Z X}$, giving \cite{Agrawal:1990jk, Buras:1998raa, Xu:2015wja}
\begin{eqnarray}
H_0 (x) & = & \epsilon_{\gamma X} D_0 (x) + \epsilon_{Z X} \tilde{D}_0 (x), \\
D_0 (x) & = & - \frac{4}{9} \ln x + \frac{(-19 x^3 + 25 x^2)}{36(x-1)^3} + \frac{x^2(5x^2 - 2x - 6)}{18(x-1)^4} \ln x, \\
\tilde{D}_0 (x) & = & - \frac{1}{s_W c_W} \Bigl [ \frac{(34x^3 - 141x^2 + 147x - 58)}{216(x-1)^3} + \frac{(-3x^4 + 18 x^3 - 27x^2 + 19 x - 4) \ln x}{36(x-1)^4} \nonumber \\
& & + c_W^2 \Bigl ( \frac{(-47x^3 + 237 x^2 - 312 x + 104)}{108(x-1)^3} + \frac{(3x^4 - 30 x^3 + 54 x^2 - 32 x + 8) \ln x}{18(x-1)^4} \Bigr ) \Bigr ]. \ \ \ \ \ \ \ \ \ \
\end{eqnarray}

\begin{figure}[h]
\centering
\subfloat[$i \mathcal{M}_a^{\rm kin.}$]
{\includegraphics[width=0.24\textwidth]{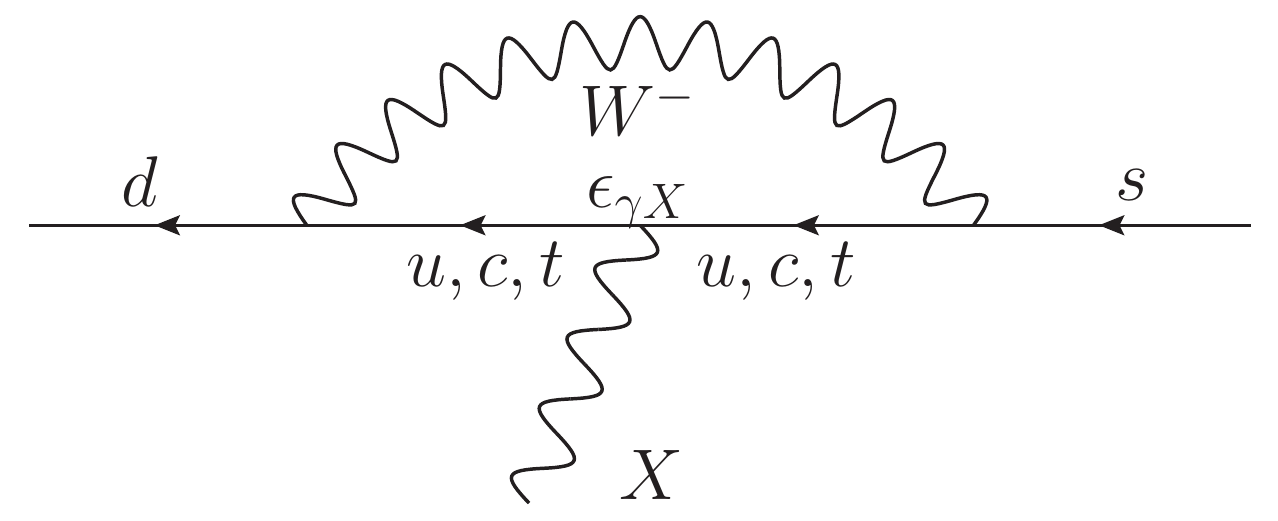}\label{fig_FCNC_kin_mix_diagram_a}} \,
\subfloat[$i \mathcal{M}_b^{\rm kin.}$]
{\includegraphics[width=0.24\textwidth]{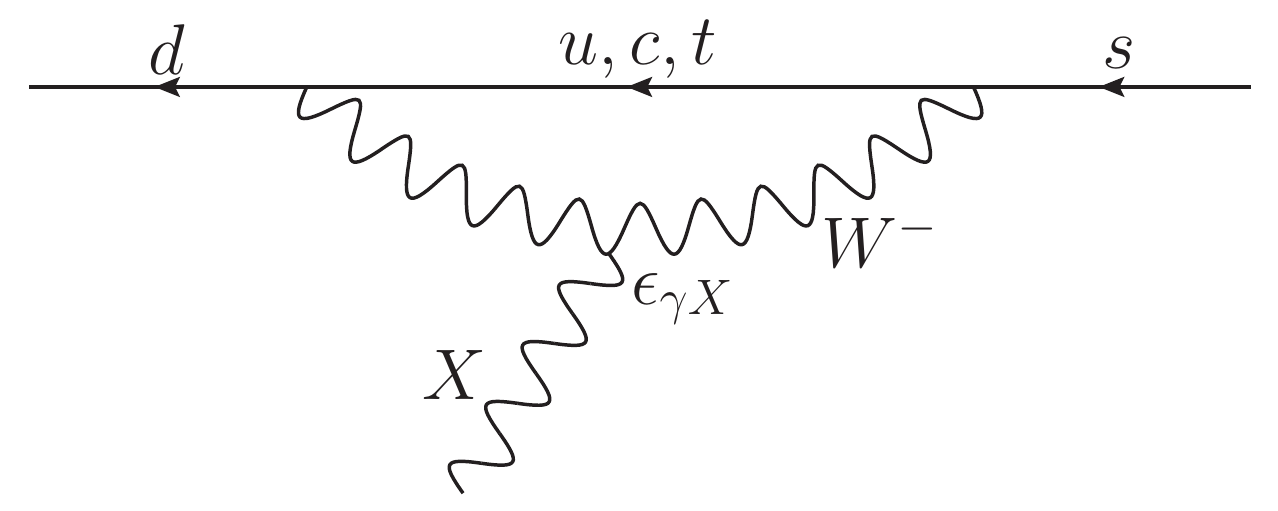}\label{fig_FCNC_kin_mix_diagram_b}} \,
\subfloat[$i \mathcal{M}_c^{\rm kin.}$]
{\includegraphics[width=0.24\textwidth]{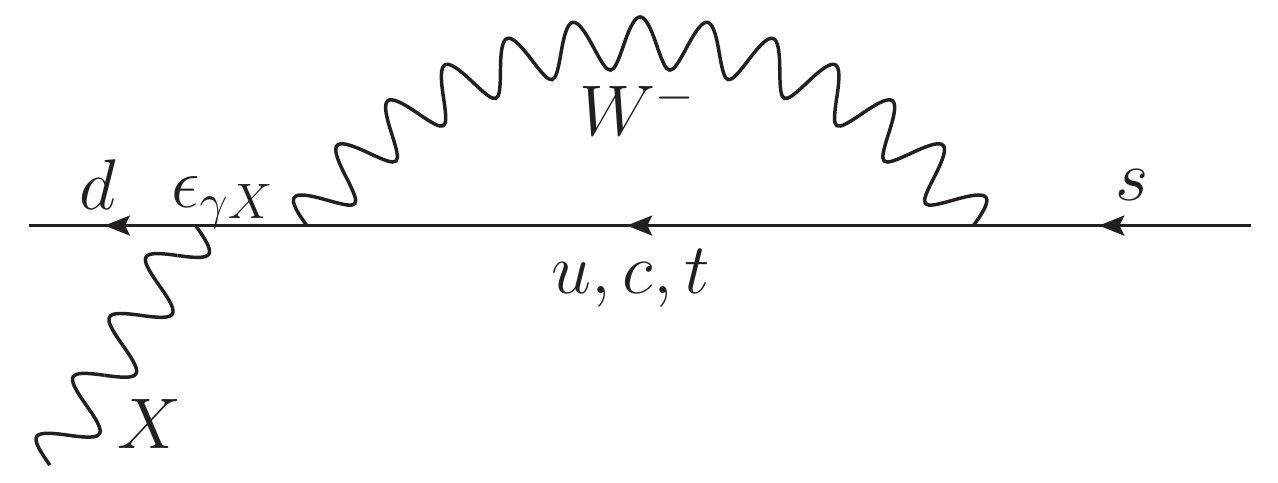}\label{fig_FCNC_kin_mix_diagram_c}} \,
\subfloat[$i \mathcal{M}_d^{\rm kin.}$]
{\includegraphics[width=0.24\textwidth]{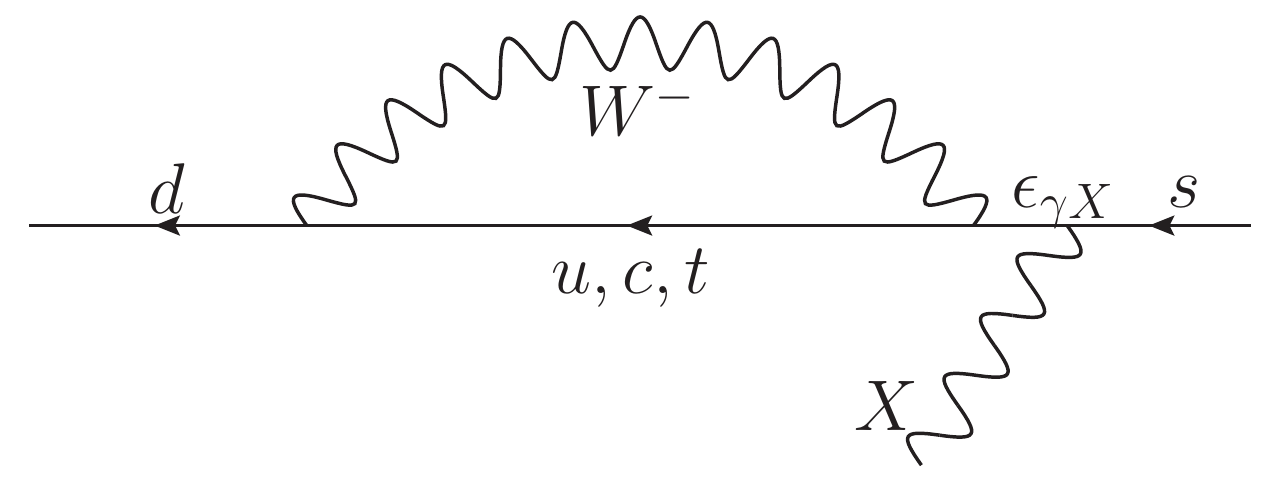}\label{fig_FCNC_kin_mix_diagram_d}} \,
  \caption{The diagrams which contribute to the loop-induced $dsX$ FCNC process only with the kinetic mixing $\epsilon_{\gamma X}$ between $X$ boson and SM photon. In $'$t Hooft-Feynman gauge, the charged goldstone boson $\phi^-$ contributions also should be included.} \label{fig_FCNC_kin_mix_diagrams}
\end{figure}

The loop function $D_0 (x)$ is determined by the sum of amplitudes $i \mathcal{M}_{a,b,c,d}^{\rm kin.}$. The diagrams for each amplitudes are shown in Fig. \ref{fig_FCNC_kin_mix_diagrams}. Due to the unbroken $U(1)_{\rm EM}$ gauge symmetry and the cancellation between the amplitudes, the resulting FCNC operator from kinetic mixing $\epsilon_{\gamma X}$ is proportional to the transverse part of the outgoing momentum, $g^{\mu \nu} q^2 - q^\mu q^\nu$.

The amplitude square for $K_L(p_1) \to \pi^0(p_2)+X(q)$ and $K^+(p_1) \to \pi^+(p_2)+X(q)$ 
are given by
\begin{eqnarray}
\frac{1}{2}|M_{K_L \to \pi^0+X}|^2 & =&  \left({\rm Im}[g^{\rm eff}_{sbX}|_{\epsilon_{\gamma X}}]\right)^2
(2\, p_1 \cdot p_2+(4\,p_1 \cdot q\, p_2 \cdot q)/m^2_X)\,, \nonumber \\
\frac{1}{2}|M_{K^+ \to \pi^++X}|^2 & =&  \vert g^{\rm eff}_{sbX}|_{\epsilon_{\gamma X}} \vert^2
(2\, p_1 \cdot p_2+(4\,p_1 \cdot q\, p_2 \cdot q)/m^2_X)\,, \nonumber \\
g^{\rm eff}_{sbX}|_{\epsilon_{\gamma X}} & \equiv & 
m^2_X \left( \sum_{i=c,t} V_{is} V_{id}^* \frac{G_F}{\sqrt{2}} \frac{e}{8\pi^2} H_0 (x_i) \right) 
\end{eqnarray}
Here we assume $\epsilon_{ZX} \simeq 0$ to simplify the discussion. 
Taking $m_X = m_{\pi^0}$, the branching ratios versus $\epsilon_{\gamma X}$ are shown in Fig.~\ref{fig:KL_piX}, after integrating the phase space and including both the $t$ and $c$ quark contributions. Here we adopt the Wolfenstein parameterization up to $\mathcal{O}(\lambda^5)$ \cite{Buras:1998raa} as
\begin{eqnarray}
V_{\rm CKM}^{\rm SM} & = & \begin{pmatrix} 
1 - \frac{\lambda^2}{2} - \frac{\lambda^4}{8} & \lambda & A \lambda^3 (\rho - i \eta) \\ 
- \lambda + \frac{1}{2} A^2 \lambda^5 [ 1 - 2(\rho + i \eta)] & 1 - \frac{\lambda^2}{2} - \frac{\lambda^4}{8} (1+ 4 A^2) & A \lambda^2 \\ 
A \lambda^3 \Bigl [ 1 - (\rho + i \eta) \left ( 1 - \frac{\lambda^2}{2} \right ) \Bigr ] \ \ \ & - A \lambda^2 + \frac{1}{2} a (1-2 \rho) \lambda^4 - i \eta A \lambda^4 \ \ \ & 1 - \frac{1}{2} A^4 \lambda^4
\end{pmatrix} + \mathcal{O}(\lambda^6) \nonumber \\
\end{eqnarray}
where $\lambda = 0.22453 \pm 0.00044$, $A=0.836\pm 0.015$, $\rho = 0.122^{+0.018}_{-0.017}$ and $\eta=0.355^{+0.012}_{-0.011}$ from the best global fit values~\cite{Tanabashi:2018oca}.

As a result, the upper bounds on the kinetic mixing from Br($K^+ \rightarrow \pi^+ X$) and the preferred value to explain KOTO excess are
\begin{eqnarray}
\epsilon_{\gamma X} & \lsim & 1.3 \times 10^{-2} \ \ \ \ \ \ \text{(E949: ${\rm Br}(K^+ \rightarrow \pi^+ X)\leq 5.6\times 10^{-8}$ for $q^2 = m_{\pi^0}^2$)}
\\
\epsilon_{\gamma X} & \simeq & 1.5 \, \ \ \ \ \ \ \ \ \ \ \ \ \ \ \ \text{(KOTO: ${\rm Br}(K_L \rightarrow \pi^0 X)= 2.1\times 10^{-9}$ for $q^2 = m_{\pi^0}^2$)} 
\end{eqnarray}

It is clear to see that the mixing should be as large as 
$\epsilon_{\gamma X} \sim \mathcal{O}(1)$, to match the required effective coupling of Eq.\,(\ref{eq:geff_KOTO}) to explain KOTO result. Note the hierarchy between the real and imaginary components of FCNC coupling $g_{dsX}^{\rm eff}$ from the charm quark contribution which is proportional to $V_{cs} V_{cd}^* D_0 (x_c)$. Avoiding constraints from the upper limit of Br($K^+ \rightarrow \pi^+ X$), which corresponds to $\epsilon_{\gamma X} \lsim 10^{-2}$, is not possible in the presence of charm quark contribution. Furthermore, invisible dark photon searches from BaBar \cite{Lees:2017lec} and NA64 \cite{Banerjee:2017hhz} exclude large kinetic mixing down to $\epsilon_{\gamma X} \lsim 10^{-3}$.

For short summary, the KOTO event excess cannot be explained by a (invisibly decaying) light gauge boson, kinematically mixed with the SM photon. Therefore, in the next two subsections, we \emph{i)} introduce heavy VLQs to enhance the coupling between $L_\mu - L_\tau$ gauge boson and SM quarks, especially for tree-level FCNC, or \emph{ii)} consider a $(L_\mu - L_\tau) + \epsilon (B_3 - L_\tau)$ gauge boson which dominantly contribute to the down-type FCNC at the loop-level.

\section{Model I: gauged $L_\mu - L_\tau$ with heavy VLQs}
\label{subsec:Model1_LmuLtau_w_VLQs}

We focus on the extension of SM gauge group by a new abelian and anomaly free $U(1)_{L_\mu-L_\tau}$ with the associated $X$ vector gauge boson~\cite{Foot:1990mn,He:1990pn,He:1991qd,Baek:2001kca}. 
As the original gauge symmetry is leptonic so that it does not allow the direct coupling to hadrons, 
the $X$ boson still can couple to the SM quark sector through the dimension-6 operators 
with cutoff $\Lambda$ at TeV scale.  When a scalar $\Phi$ carrying $+1$ charge under $U(1)_{L_\mu-L_\tau}$~\cite{Altmannshofer:2014cfa} is introduced,  the relevant dimension-6 operators are explicitly given as
$$
\mathcal{L}_{\text{dim-6}}=(-i(D_\alpha \Phi)^*\Phi+i\Phi^*(D_\alpha \Phi))
\left[ \frac{\lambda^q_{ij}}{\Lambda^2}(\bar{q}^i_L \gamma^\alpha q^j_L)+
       \frac{\lambda^d_{ij}}{\Lambda^2}(\bar{d}^i_R \gamma^\alpha d^j_R)+
       \frac{\lambda^u_{ij}}{\Lambda^2}(\bar{u}^i_R \gamma^\alpha u^j_R) \right]\,,
$$
where $q_L=(u_L,d_L)$, $d_R$, and $u_R$ are the $SU(2)_L$ doublet and singlet quarks 
with flavour index $i,j$.
In general, the coupling $\lambda^{q,d,u}_{i,j}$ are $3\times 3$ complex matrices,
which potentially violate flavour and CP symmetries.
After $\Phi$ gets a VEV $\langle \Phi \rangle=v_\Phi/\sqrt{2}$, 
the $U(1)_{L_\mu-L_\tau}$ is spontaneously broken,
and then the hadronic current violating flavour symmetry is generated
\begin{eqnarray}
\label{eq:had_current}
J_X^{\mu\text{(had)}} & = & \mathbb{R}_{ij}^{(d)} \bar{d}_i \gamma^\mu P_R d_j + \mathbb{L}_{ij}^{(d)} \bar{d}_i \gamma^\mu P_L d_j + \mathbb{R}_{ij}^{(u)} \bar{u}_i \gamma^\mu P_R u_j + \mathbb{L}_{ij}^{(u)} \bar{u}_i \gamma^\mu P_L u_j\,.
\end{eqnarray}
Explicit forms of $ \mathbb{R}_{ij}^{(u,d)}, \mathbb{L}_{ij}^{(u,d)} $ are given in Eq.\,(\ref{eq:FCNC}).
At the same time, the $X$ boson obtain a mass $m_X=g_X v_\Phi$, 
where $g_X$ is the $U(1)_{L_\mu-L_\tau}$ gauge coupling. After all, the effective action is given as
\begin{eqnarray}
\mathcal{L} & \supset & \mathcal{L}_{\rm SM} - \frac{1}{4} X_{\mu\nu} X^{\mu\nu} + \frac{m_X^2}{2} X_\mu X^\mu - g_X X_\mu J^{\mu(\rm lep)}_X - g_X X_\mu J^{\mu(\rm had)}_X,
\end{eqnarray}
where $X_{\mu\nu}\equiv\partial_\mu X_\nu -\partial_\nu X_\mu$
is the field strength tensor of $X_\mu$. The leptonic current corresponding to $L_\mu-L_\tau$ is
\begin{eqnarray}
\label{eq:jmu}
J_X^{\mu (\rm lep)} & = & \left ( \bar{\ell}_{2 L} \gamma^\mu \ell_{2 L} + \bar{\mu}_R \gamma^\mu \mu_R \right ) - \left ( \bar{\ell}_{3 L} \gamma^\mu \ell_{3 L} + \bar{\tau}_R \gamma^\mu \tau_R \right )
\end{eqnarray}
where $g_X\simeq 5\times 10^{-4}$ and $m_X\leq 2m_\mu$ is still allowed region for the $(g-2)_\mu$ anomaly~\cite{Baek:2001kca,Pospelov:2008zw}, and $\ell_{2L} = (\nu_\mu \ \mu)_L^T$ and $\ell_{3L} = (\nu_\tau \  \tau)_L^T$ are the second and third generation lepton doublets in the SM, respectively. To explain the KOTO events excess, additional $X$ boson couplings to $J^{\mu(\text{had})}_X$ will be generated by introducing VLQs at TeV scale, which mix with SM quarks.

\begin{figure}[h]
\centering
\subfloat[down-type FCNC]
{\includegraphics[width=0.43\textwidth]{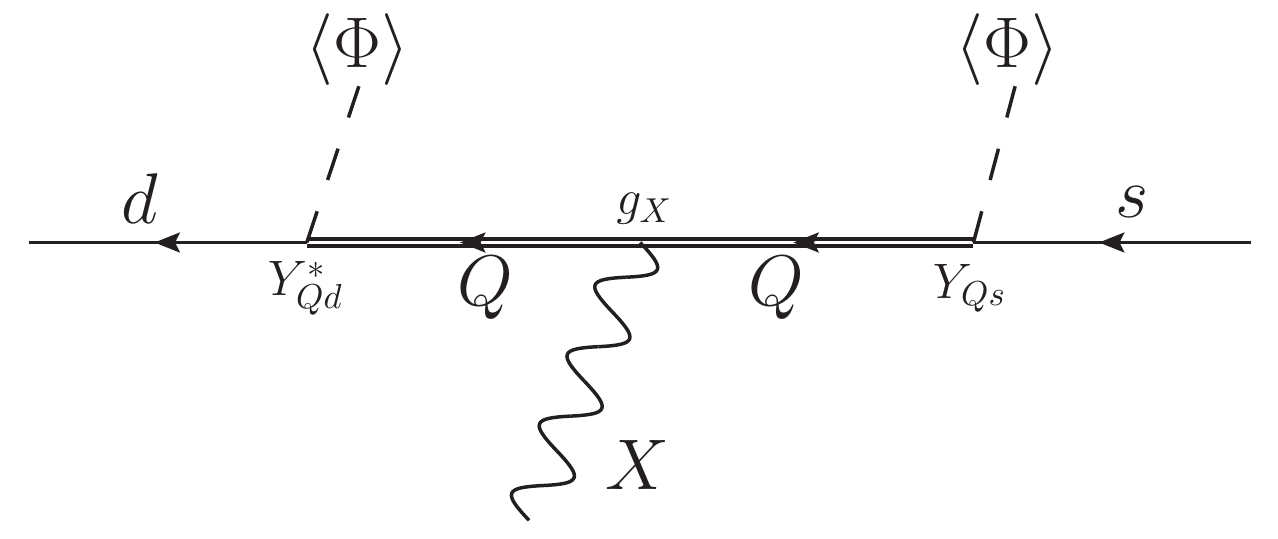}\label{fig_FCNC_VLQ_Q_dsX_diagram_a}} \,
\subfloat[up-type FCNC]
{\includegraphics[width=0.43\textwidth]{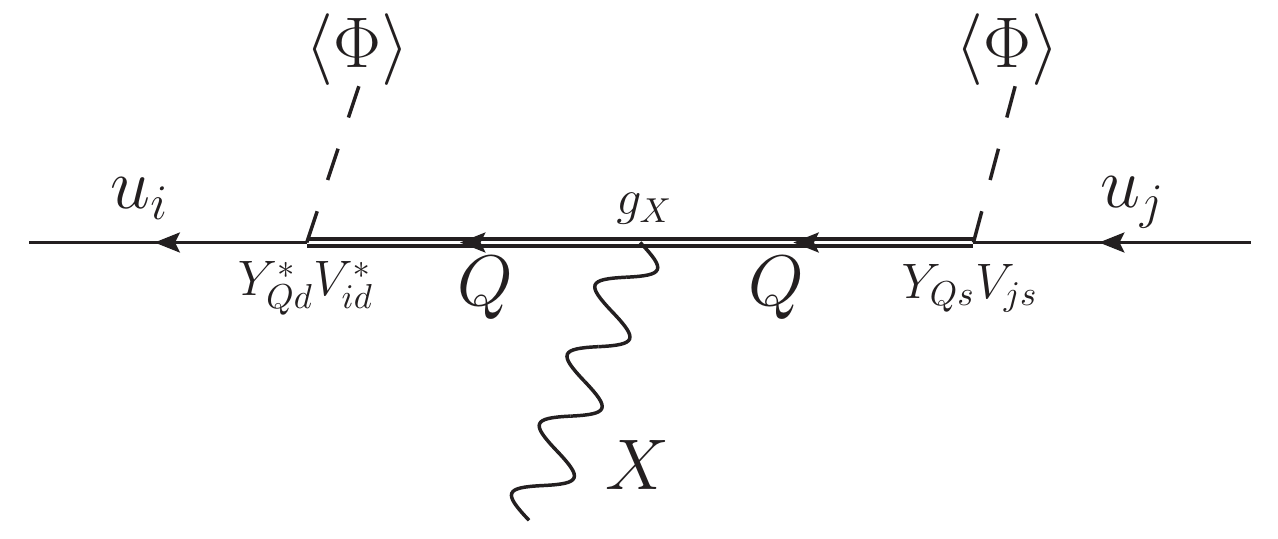}\label{fig_FCNC_VLQ_Q_dsX_diagram_b}} \,
  \caption{The diagrams which contribute to $dsX$ FCNC process with $X$ gauge boson coupled to a heavy vector-like quark $Q$ (model I).}\label{fig_FCNC_VLQ_Q_dsX_diagrams}
\end{figure}

The general expression for the hadronic current induced from VLQs, 
which carry $U(1)_{L_\mu-L_\tau}$ charges 
and couple to SM quark sector through new scalar $\Phi$. 
The model has been previously suggested in Ref.\,\cite{Altmannshofer:2014cfa} to explain the lepton universality violation (LUV) in rare $B$ meson decay $B \rightarrow K^* l^- l^+$ $(l=e,\mu)$ and has been applied to $K_L^0 \rightarrow \pi^0 + (\text{invisible})$ with enhanced coupling to top quark \cite{Fuyuto:2014cya}.
We follow Ref.\,\cite{Altmannshofer:2014cfa} and introduce VLQs with 
the gauge charges $(SU(3)_c, SU(2)_L)_{(Y, Q')}$ of the interaction eigenstates are assigned as
\begin{eqnarray}
Q_L & = & 
\left(\begin{array}{c} 
U_L \\ D_L 
\end{array}\right) 
= ({\bf 3}, {\bf 2})_{(+ \frac{1}{6}, +1)}\ , \quad
\tilde{Q}_R =  
\left(\begin{array}{c}
\tilde{U}_R \\ \tilde{D}_R 
\end{array}\right)
= ({\bf 3}, {\bf 2})_{(+ \frac{1}{6}, +1)}\ , \nonumber \\ 
\tilde{U}_L & = & ({\bf 3}, {\bf 1})_{(+ \frac{2}{3}, -1)}\ , \quad \quad \quad \quad \quad \quad
U_R  =  ({\bf 3}, {\bf 1})_{(+ \frac{2}{3}, -1)}, \nonumber \\
\tilde{D}_L & = & ({\bf 3}, {\bf 1})_{(- \frac{1}{3}, -1)}\ , \quad \quad \quad \quad \quad \quad
D_R  =  ({\bf 3}, {\bf 1})_{(- \frac{1}{3}, -1)}\ ,
\end{eqnarray}
where $Y$ and $Q'$ are SM hypercharge and $U(1)_{L_\mu-L_\tau}$ charge, respectively.
Then the Yukawa interactions between VLQs and SM quarks are written as
\begin{eqnarray}
\mathcal{L}_{\rm mix} & = & 
\Phi \bar{\tilde{D}}_R (Y_{Qb}b_L+ Y_{Qs}s_L+Y_{Qd}d_L)
+\Phi \bar{\tilde{U}}_R (Y_{Qt}t_L+ Y_{Qc}c_L+Y_{Qu}u_L)\, \nonumber \\
& & + \Phi^\dagger \bar{\tilde{U}}_L (Y_{Ut}t_R+ Y_{Uc}c_R+Y_{Uu}u_R)
+\Phi^\dagger \bar{\tilde{D}}_L (Y_{Db}b_R+ Y_{Ds}s_R+Y_{Dd}d_R)+{\rm h.c} \,, \ \ \ \ \ \ \ \ \
\end{eqnarray}
that will induce the mixing between VLQs and SM quarks.
In order to maintain the electroweak invariant, they shall satisfy the relation
\begin{eqnarray}
\label{eq:up_down}
(Y_{Qu},Y_{Qc},Y_{Qt})^{T} = (V^{\rm SM}_{\rm CKM})^*\,(Y_{Qd},Y_{Qs}, Y_{Qb})^T\,.
\end{eqnarray}
After $\Phi$ gets VEV, these Yukawa interactions contribute to the off diagonal elements of the up-type and down-type quark mass matrices
\begin{eqnarray}
\label{eq:mass_matrix}
\mathcal{M}_u= \left(
\begin{array}{ccccc}
m_u & 0 & 0 & 0 & \frac{Y_{Qu}\, v_\Phi}{\sqrt{2}} \\ 
0 & m_c & 0 & 0 & \frac{Y_{Qc}\, v_\Phi}{\sqrt{2}} \\  
0 & 0 & m_t & 0 & \frac{Y_{Qt}\, v_\Phi}{\sqrt{2}} \\  
\frac{Y_{Uu}\, v_\Phi}{\sqrt{2}} & \frac{Y_{Uc}\, v_\Phi}{\sqrt{2}} & \frac{Y_{Ut}\, v_\Phi}{\sqrt{2}}   & m_U & 0  \\  
0 & 0 & 0   & 0   & m_Q
\end{array}\right)\,,
\quad
\mathcal{M}_d= \left(
\begin{array}{ccccc}
m_d & 0 & 0 & 0 & \frac{Y_{Qd}\, v_\Phi}{\sqrt{2}} \\ 
0 & m_s & 0 & 0 & \frac{Y_{Qs}\, v_\Phi}{\sqrt{2}} \\  
0 & 0 & m_b & 0 & \frac{Y_{Qb}\, v_\Phi}{\sqrt{2}} \\  
\frac{Y_{Dd}\, v_\Phi}{\sqrt{2}} & \frac{Y_{Ds}\, v_\Phi}{\sqrt{2}} & \frac{Y_{Db}\, v_\Phi}{\sqrt{2}}   & m_D & 0  \\  
0 & 0 & 0   & 0   & m_Q
\end{array}\right)\,, \nonumber \\
\end{eqnarray}
where the masses for VLQs come from
\begin{eqnarray}
\mathcal{L}\supset m_Q \bar{Q}_L \tilde{Q}_R+m_D \bar{\tilde{D}}_L {D_R}
+m_U \bar{\tilde{U}}_L {U_R}+\text{h.c.}\,.
\end{eqnarray}
They can be diagonalized by performing the bi-unitary transformation: 
\begin{eqnarray}
{\cal U}_L {\cal M}_u {\cal U}^\dagger_R &=& {\cal M}^{\rm diag}_u\,, \quad
\text{or} \quad
{\cal U}_L {\cal M}_u {\cal M}_u^\dagger {\cal U}^\dagger_L =
{\cal U}_R {\cal M}_u^\dagger {\cal M}_u {\cal U}^\dagger_R =
({\cal M}^{\rm diag}_u)^2 \,, \nonumber \\
{\cal D}_L {\cal M}_d {\cal D}^\dagger_R &=& {\cal M}^{\rm diag}_d\,, \quad
\text{or} \quad
{\cal D}_L {\cal M}_d {\cal M}_d^\dagger {\cal D}^\dagger_L =
{\cal D}_R {\cal M}_d^\dagger {\cal M}_d {\cal D}^\dagger_R =
({\cal M}^{\rm diag}_d)^2 \,,
\end{eqnarray}
and induce the FCNC interactions for Eq./,(\ref{eq:had_current})
\begin{eqnarray}
\label{eq:FCNC}
\mathbb{L}^{(d)}_{ij}=\frac{Y_{Qi}Y^*_{Qj}v^2_\Phi}{2m^2_Q}\,,~
\mathbb{R}^{(d)}_{ij}=-\frac{Y_{Di}Y^*_{Dj}v^2_\Phi}{2m^2_D}\,,~
\mathbb{L}^{(u)}_{ij}=
\left(V^{\rm SM}_{\rm CKM}\mathbb{L}^{(d)}V^{\rm SM \dagger}_{\rm CKM} \right)_{ij}\,,~
\mathbb{R}^{(u)}_{ij}=-\frac{Y_{Ui}Y^*_{Uj}v^2_\Phi}{2m^2_U}\,, \nonumber \\
\end{eqnarray} 
which are illustrated in Fig.~\ref{fig_FCNC_VLQ_Q_dsX_diagrams}. Here we only keep the leading terms of order $v^2_\Phi/m^2_{Q,U,D}$. The left-handed FCNC, $g_X\mathbb{L}^{(d)}_{sd}=g_X(Y_{Qs}Y^*_{Qd}v^2_\Phi/(2m^2_Q))$ 
comes from $\tilde{Q}_R$, 
is the most relevant tree-level coupling for $K_L \to \pi^0 X$.

Furthermore, the CKM unitarity within $3 \times 3$ SM quark block 
will be violated due to the extension of quark sector. 
In fact, the CKM matrix is extending to $5 \times 5$ and relates to SM CKM matrix as
\begin{eqnarray}
{\bf V}_{\rm CKM}=
\mathcal{U}_L  \left( \begin{array}{ccc}
      {\bf V}^{\rm SM}_{\rm CKM}  & 0 & 0\\
      0  & 0  & 0 \\
      0  & 0  & 1 \end{array} \right) \mathcal{D}^\dagger_L\,,
\end{eqnarray}
where the zero diagonal element 
is from $SU(2)_L$ singlet VLQs $\tilde{U}_L$ and $\tilde{D}_L$. 
The unitarity condition still hold in the $5 \times 5$ CKM matrix, but it would be violated in the $3 \times 3$ block of CKM, that can be tested by current precision measurements of CKM elements, for example the deviation from unity should be less than $\mathcal{O}(10^{-3})$ under current measurements on 1st row of CKM, i.e. $|V_{ud}|^2+|V_{cd}|^2+|V_{td}|^2=0.9983(4)$~\cite{Seng:2018yzq}.

\subsection{Explanation of KOTO events}
\label{subsec:Width_Model1}

We have added six VLQs: $Q_L$, $\tilde{Q}_R$, $\tilde{U}_L$, $U_R$, $\tilde{D}_L$, and $D_R$. Only $\tilde{Q}_R$ gives most relevant contribution to $K\to \pi X$, which involves the left-handed down quarks mixing among 1st and 2nd generations. From Eq.\,(\ref{eq:mass_matrix}), we can see $\tilde{Q}_R$ generates non-zero off-diagonal elements on the upper-right corner, meanwhile, $\tilde{U}_L$ and $\tilde{D}_L$ contribute on the lower-left corner. According to these patterns, as it was expressed in Eq.\,(\ref{eq:FCNC}), that $\tilde{Q}_R$ induce the larger left-handed quarks mixing, while  $\tilde{U}_L$ and $\tilde{D}_L$ induce the larger right-handed mixing. As a consequence, introducing only $\tilde{Q}_R$ is the efficient way to enhance the $K\to \pi X$. They give the tree-level FCNC effective interactions as
\begin{eqnarray}
\mathcal{L}_{\rm eff} & \supset &
\frac{g_X(Y_{Qd}Y^*_{Qs})v^2_\Phi}{2m^2_Q}
[\bar{d}_L\gamma^\mu s_L]X_\mu
+\frac{g_X(Y_{Qs}Y^*_{Qb})v^2_\Phi}{2m^2_Q}
[\bar{s}_L\gamma^\mu b_L]X_\mu  \nonumber \\
& & 
+ \frac{g_X(Y_{Qu}Y^*_{Qc})v^2_\Phi}{2m^2_Q}
[\bar{u}_L\gamma^\mu c_L]X_\mu
+\frac{g_X(Y_{Qc}Y^*_{Qt})v^2_\Phi}{2m^2_Q}
[\bar{c}_L\gamma^\mu t_L]X_\mu
+ {\rm h.c.} \nonumber \\
& \equiv & 
g^{\rm eff}_{dsX}|_{\rm VLQ}\,
[\bar{d}_L\gamma^\mu s_L]X_\mu
+g^{\rm eff}_{sbX}|_{\rm VLQ}\,
[\bar{s}_L\gamma^\mu b_L]X_\mu \nonumber \\
& & 
+ g^{\rm eff}_{ucX}|_{\rm VLQ}\,
[\bar{u}_L\gamma^\mu c_L]X_\mu
+g^{\rm eff}_{ctX}|_{\rm VLQ}\,
[\bar{c}_L\gamma^\mu t_L]X_\mu
+ {\rm h.c.}\,.
\end{eqnarray}
However, $\tilde{Q}_R$ also induces non-trivial FCNC for the up-type quarks due to the relation of Eq.\,(\ref{eq:up_down}),
but, the FCNC constraints among up-quark sector are not as stringent as 
the down-quark sector.
To explain the KOTO event excess, the effective coupling $g^{\rm eff}_{dsX}|_{\rm VLQ}$ shall satisfy Eq.\,(\ref{eq:geff_KOTO}) and gives
\begin{eqnarray}
\label{eq:KOTO_perfer}
\frac{{\rm Im}(Y_{Qs}Y^*_{Qd})}{2m^2_Q}\simeq \frac{3.43\times 10^{-14}}{\rm GeV^2}\,,
\end{eqnarray}
by fixing $g_X=5\times 10^{-4}$ and $v_\Phi=260$ GeV 
to give $m_X= g_X v_\Phi\simeq 135$ MeV close to neutral pion mass. 
The Yukawa coupling strengths are estimated to be $Y_{Qs}\simeq Y_{Qd}\simeq 5.2\times 10^{-4}$, if we choose $m_Q\simeq 2$ TeV, which is heavy enough to satisfy all the current mass lower bound from the VLQs direct searches at the LHC \cite{Aaboud:2018pii, Sirunyan:2018qau, Sirunyan:2019sza, Sirunyan:2017pks, Aaboud:2018ifs, ATLAS:2018qxs}.

\subsection{constraints}
\label{subsec:constraint_model1}

\subsubsection{$K^0-\bar{K}^0$ mixing}

The CP violation in Kaon mixing process might put strong bound on the FCNC between the 1st and 2nd generations in down-quark sector. In terms of six-dimensional operator
\begin{eqnarray}
\Delta \mathcal{L}_{\Delta F=2}^{(sd)} & = & \frac{1}{\Lambda_{ds}^2} (\bar{d}_L \gamma^\mu s_L) (\bar{d}_L \gamma_\mu s_L),
\end{eqnarray}
the upper bound of the FCNC coupling $(g_{dsX}^{\rm eff})^2$ can be translated into the lower bound on the scale $\Lambda_{ds}$. The lower bound on $\Lambda_{ds}$ comes from the experimental constraints on the mass difference $\Delta m_K$ and the mixing coefficient $\epsilon_K$. We quote limits from \cite{Isidori:2010kg}
\begin{eqnarray}
| \text{Re} \ ( \Lambda_{ds}^{-2}) | & \ \leq \ & 9.0 \times 10^{-13} \text{ GeV}^{-2} , \label{KaonMixingRe} \\
| \text{Im} \ ( \Lambda_{ds}^{-2}) | & \ \leq \ & 3.4 \times 10^{-15} \text{ GeV}^{-2} \label{KaonMixingIm}
\end{eqnarray}
as the constraint from Kaon mixing in this work.

Nevertheless, the FCNC coupling induced by heavy VLQ can contribute to the FCNC operator as
\begin{eqnarray}
\Delta \mathcal{L}_{\Delta F=2}^{(sd)} & = & \frac{(g_{sdX}^{\rm eff})^2}{m_K^2 - m_X^2} (\bar{d}_L \gamma^\mu s_L) (\bar{d}_L \gamma_\mu s_L) = \frac{1}{m_K^2 - m_X^2} \Bigl (g_X \frac{Y_{Qs} Y_{Qd}^* v_\Phi^2}{2m_Q^2} \Bigr )^2 (\bar{d}_L \gamma^\mu s_L) (\bar{d}_L \gamma_\mu s_L) \nonumber \\
\end{eqnarray}
and it gives the upper bounds
\begin{eqnarray}
| \text{Re} \{ (g_{sdX}^{\rm eff})^2 \} | \simeq | (\text{Re} \ g_{sdX}^{\rm eff})^2 - (\text{Im} \ g_{sdX}^{\rm eff})^2 | & \ \leq \ & 2.06 \times 10^{-13}, \\
| \text{Im} \{ (g_{sdX}^{\rm eff})^2 \} | \simeq | 2(\text{Re} \ g_{sdX}^{\rm eff})(\text{Im} \ g_{sdX}^{\rm eff}) | & \ \leq \ & 7.80 \times 10^{-16}
\end{eqnarray}
at $m_X = m_{\pi^0}$. The KOTO desired (and allowed by $K^+ \rightarrow \pi^+ X$ branching ratio measurement) region satisfies Kaon mixing constraints, with large difference of the order of magnitude. If we assume $g_X \simeq 5 \times 10^{-4}$ and $v_\Phi \simeq 260$ GeV, they give
\begin{eqnarray}
\label{eq:kkmixing_schannel}
\frac{| \text{Re} \{(Y_{Qs} Y_{Qd}^*)^2\} |}{2m_Q^4} & \leq & 3.61 \times 10^{-16} \text{ GeV}^{-4}, \\
\frac{| \text{Im} \{(Y_{Qs} Y_{Qd}^*)^2\} |}{2m_Q^4} & \leq & 1.40 \times 10^{-18} \text{ GeV}^{-4},
\end{eqnarray}
implying $| Y_{Qd,Qs} |/m_Q \lsim \mathcal{O}(10^{-4}-10^{-5}) \text{ GeV}^{-1}$. The KOTO desired value is expected to have $| Y_{Qd,Qs} |/m_Q \sim \mathcal{O}(10^{-6}) \text{ GeV}^{-1}$, and still comfortably survives.

 For chirality-flipping operator, the new physics bound becomes slightly stronger, but the KOTO desired values are not excluded. Even in the presence of both $\tilde{Q}_R$ and $\tilde{D}_L$ (and mixing to the SM $s$ and $d$ quarks), we find that the Kaon mixing constraint is not sensitive to our bulk part parameters. The Kaon mixing constraint can be translated into the bound on flavour-changing couplings to both left-handed and right-handed quarks through the effective operator $(\bar{s}_Rd_L)(\bar{s}_L d_R)$ from box diagrams~\cite{Altmannshofer:2014cfa,Isidori:2010kg}
\begin{equation}
\frac{| {\rm Im}(Y_{Qs}Y^*_{Qd}Y_{Ds}Y^*_{Dd}) |}{2m^2_Q m^2_D}
\lsim \frac{1.48\times 10^{-22}}{\rm GeV^4}\,.\label{eq:KKmixing_box}
\end{equation}
Compared to KOTO preferred parameter region, i.e. $v_\Phi\simeq 260$ GeV, $m_{Q,D}\simeq 2$ TeV, $Y_{Qs,Qd}\simeq 5.2\times 10^{-4}$, similar values of $Y_{Qs,Qd}\simeq Y_{Qs,Qd}$ are required to be electroweak invariant, 
then they give
\begin{equation}
\frac{| {\rm Im}(Y_{Qs}Y^*_{Qd}Y_{Ds}Y^*_{Dd}) |}{2m^2_Q m^2_D}
\simeq \frac{2.35\times 10^{-27}}{\rm GeV^4},
\end{equation}
which still satisfies the Kaon mixing constraints 
from Eq.\,(\ref{eq:kkmixing_schannel}) and Eq.\,(\ref{eq:KKmixing_box}).

Since we impose the mixings between the heavy vector-like quark $\tilde{Q}_R$ and the left-handed SM quarks $s_L$ and $d_L$, it naturally provides up-type quark interactions including flavour violating components as
\begin{eqnarray}
\mathbb{L}^{(u)} & = & V_{\rm CKM}^{\rm SM} \cdot \mathbb{L}^{(d)} \cdot V_{\rm CKM}^{\rm SM\dagger} =  \frac{v_\Phi^2}{2m_Q^2} \cdot V_{\rm CKM}^{\rm SM} \begin{pmatrix}
| Y_{Qd} |^2 & Y_{Qd} Y_{Qs}^* & 0 \\
Y_{Qs} Y_{Qd}^* & | Y_{Qs} |^2 & 0 \\
0 & 0 & 0
\end{pmatrix} V_{\rm CKM}^{\rm SM\dagger}
\end{eqnarray}
due to the $SU(2)_L$ gauge invariance. If we assume $\mathcal{O}(10^{-3})$ of real and imaginar components of yukawa couplings $Y_{Qd}$ and $Y_{Qs}$, then we get very tiny couplings for $g_{ucX}^{\rm eff} \sim \mathcal{O}(10^{-17})$. It is obviously safe from current upper bounds obtained by $D$ meson mixing constraints.

\subsubsection{$K_L \to \mu^+ \mu^-$} \label{subsubsec:KLtomumu_VLQ}

The CP-conserving Kaon rare decay $K_L \to \mu^+ \mu^-$ 
has similar short-distance part contribution
to $K_L \to \pi \nu \bar{\nu}$, from $Z$-penguins and box diagrams.
However, important long-distance contributions from two-photon intermediate state
are difficult to precisely be calculated and separated from short-distance part.
Therefore, here we just require the additional contributions from our model of 
${\rm Br}(K_L \to \mu^+ \mu^-)$
do not exceed the current experimental observation~\cite{Tanabashi:2018oca}
\begin{eqnarray}
{\rm Br}(K_L \to \mu^+ \mu^-)_{\rm EXP}=(6.84\pm 0.11)\times 10^{-9}\,.
\end{eqnarray}

For the short-distance part, the effective Hamiltonian from SM and VLQ are~\cite{Buras:1997fb}
\begin{eqnarray}
\mathcal{H}^{\rm SM}_{\rm eff}&=& -\frac{G_F}{\sqrt{2}}\frac{\alpha_{\rm EM}}{2\pi \sin^2 \theta_w}
\left( V^*_{cs}V_{cd}Y_{\rm NL}+V^*_{ts}V_{td}Y(x_t) \right)
[\bar{s}\gamma^{\mu}P_L d][\bar{\mu}\gamma_{\mu}P_L \mu]+{\rm h.c}\,, \nonumber \\
\mathcal{H}^{\rm VLQ}_{\rm eff}&=&
-g^2_x v^2_{\Phi} \frac{Y_{Qs}Y^*_{Qd}}{2m^2_Q}
\left(\frac{1}{m^2_{K_L}-m^2_X} \right)
[\bar{s}\gamma^{\mu}P_L d][\bar{\mu}\gamma_{\mu}P_L \mu]+{\rm h.c}\,,
\end{eqnarray}
where $\alpha_{\rm EM}\equiv \frac{e^2}{4\pi}$ and the loop functions
\begin{eqnarray}
Y(x_t)&=& Y_0(x_t)+\frac{\alpha_{\rm EM}}{4\pi}Y_1(x_t)\simeq \eta_Y Y_0(x_t)\simeq 1.062\,, \nonumber \\
Y_0(x) &=& \frac{x}{8}\left[ \frac{x-4}{x-1}+\frac{3x}{(x-1)^2}\ln(x) \right]      \,,
\end{eqnarray}
with $x_t\equiv \frac{m^2_t}{m^2_W}$, and $\eta_Y=1.026\pm 0.006$.
Then the branching ratio is
\begin{eqnarray}
{\rm Br}(K_L \to \mu^+\mu^-)&=&\kappa_\mu 
\left[ \frac{{\rm Re}\lambda_c}{\lambda}P_0(Y)+
       \frac{{\rm Re}\lambda_t}{\lambda^5}Y(x_t)
       +g^2_x v^2_{\Phi} \frac{{\rm Re}(Y_{Qs}Y^*_{Qd})}{2m^2_Q}
\left(\frac{1}{m^2_{K_L}-m^2_X} \right)
\frac{2\sqrt{2}\pi \sin^2\theta_w}{G_F {\alpha_{\rm EM}} \lambda^5} \right]^2\,, 
\nonumber \\
\end{eqnarray}
where the first two terms in the square bracket are from SM short-distance part, and the third one comes from VLQs contribution. 
Here we defined
\begin{eqnarray}
\kappa_\mu &=& \frac{\alpha^2_{\rm EM} {\rm Br}(K^+ \to \mu^+ \nu)}{\pi^2 \sin^4 \theta_w }
\frac{\tau_{K_L}}{\tau_{K^+}}\lambda^8 = 1.68\times 10^{-9}
\,, 
\end{eqnarray}
where $P_0(Y)=Y_{\rm NL}/\lambda^4\simeq 0.138$ and $\lambda\equiv V_{us}=0.22453\pm 0.00044$. One obtains
\begin{eqnarray}
\lambda_t\equiv && V_{td} V^*_{ts}=(-3.41+1.45i)\times 10^{-4}  \,, \nonumber \\
\lambda_c\equiv && V_{cd} V^*_{cs}=-0.218- 1.45\times 10^{-4}i \,.
\end{eqnarray}
By insert these values, we obtained SM short-distance contribution
$$
{\rm Br}(K_L \to \mu^+\mu^-)_{\rm SM}=9.929\times 10^{-10}\,.
$$
Combining the VLQ and SM contribution and using KOTO preferred region from Eq.\,(\ref{eq:KOTO_perfer})
it gives
$
{\rm Br}(K_L \to \mu^+\mu^-)=9.931\times 10^{-10}\,,
$
which does not modify much.
Under the preferred parameter values for KOTO event excess,
VLQs contribution to ${\rm Br}(K_L \to \mu^+\mu^-)$ is less than $\mathcal{O}(10^{-12})$,
which is two orders of magnitude below the current experimental sensitivity.

\subsubsection{CKM unitarity}

Before considering the SM quarks mixing with VLQs, 
we assume that the 3 by 3 block of quark mass matrix corresponding to SM is diagonalized, 
as shown in Eq.\,(\ref{eq:mass_matrix}).
Hence the 3 by 3 block of SM CKM matrix satisfies unitarity. 

After SM quarks mixing with VLQs, 
the couplings with $W$ boson are modified as
\begin{eqnarray}
  - {\cal L} &\supset &
  g_W (\overline{u},\overline{c},\overline{t},\overline{\tilde{U}},\overline{U})_L \gamma^\mu W^+ 
  \left( \begin{array}{ccc}
      {\bf V}^{\rm SM}_{\rm CKM}  & 0 & 0\\
      0  & 0 & 0 \\
      0  & 0 & 1   \end{array} \right) 
\left(\begin{array}{c} d \\ s \\ b \\ \tilde{D} \\ D \end{array}\right)_L  \nonumber \\
& = &
  g_W (\overline{u},\overline{c},\overline{t},\overline{\tilde{U}},\overline{U})^m_L \gamma^\mu W^+ 
\, \mathcal{U}_L  \left( \begin{array}{ccc}
      {\bf V}^{\rm SM}_{\rm CKM}  & 0 & 0\\
      0  & 0  & 0 \\
      0  & 0  & 1 \end{array} \right) \mathcal{D}^\dagger_L
\left(\begin{array}{c} d \\ s \\ b \\ \tilde{D} \\ D \end{array}\right)^m_L\,,
\end{eqnarray}
where $SU(2)$ singlet $\tilde{D}_L$ and $\tilde{U}_L$ do not couple to $W$ boson.
And then the CKM is modified accordingly
\begin{eqnarray}
{\bf V}_{\rm CKM}=
\mathcal{U}_L  \left( \begin{array}{ccc}
      {\bf V}^{\rm SM}_{\rm CKM}  & 0 & 0\\
      0  & 0  & 0 \\
      0  & 0  & 1 \end{array} \right) \mathcal{D}^\dagger_L\,.
\end{eqnarray}
The first-row of CKM $|V_{ud}|^2+|V_{us}|^2+|V_{ub}|^2\simeq 1$ are known with highest precision and good agreement with unitarity. According to recent calculation of inner radiative correction with reduced hadronic uncertainty, the updated value of $|V_{ud}|=0.97366(15)$ has been obtained \cite{Seng:2018yzq}.

The preferred values of input parameter to explain KOTO events are $Y_{Qd}=Y_{Qs}=5.2\times 10^{-4}$, $Y_{Qb}=0$, $v_\Phi=260$ GeV, and $m_{Q,D}\simeq 2$ TeV. It gives the mixing angle between VLQs $Q_L$ and $d$ quark of 
\begin{equation}
\frac{Y_{Qd}\,v_\Phi}{\sqrt{2} m_Q}\simeq 0.48\times 10^{-4}\,.
\end{equation} 
Therefore, the VLQs modifications of the SM corresponding CKM matrix is of order $\mathcal{O}(10^{-4})$, it is still compatible with the present observational precision of $|V_{ud}|$.

\section{Model II: gauged $(L_\mu - L_\tau) + \epsilon (B_3 - L_\tau)$ in the presence of RH$\nu$}
\label{subsec:Model2_LmuLtau_B3Ltau_w_RHnu}

In the presence of (at least) two species of heavy right-handed neutrinos $N_{2,3}$, we can consider a possible anomaly-free extension of the gauge group $G = G_{\rm SM} \otimes U(1)_{L_\mu - L_\tau} \otimes U(1)_{B_3 - L_\tau}$ as \cite{Foot:1990mn, He:1990pn, He:1991qd, Alonso:2017uky, Babu:2017olk}
\begin{eqnarray}
\mathcal{L} & \supset & \mathcal{L}_{\rm SM} - \frac{1}{4} \sum_{i=1,2} \hat{X}_{i \mu \nu} \hat{X}_i^{\mu \nu} - \frac{\epsilon_{12}}{2} \hat{X}_{1\mu\nu} \hat{X}_2^{\mu\nu} + \frac{1}{2} \sum_{i=1,2} \hat{M}_i^2 \hat{X}_{i \mu} \hat{X}_i^\mu + \delta \hat{M}_{12}^2 \hat{X}_{1\mu} \hat{X}_2^\mu \nonumber \\
& & - \hat{g}_{X_1} J_{L_\mu - L_\tau}^\mu \hat{X}_{1\mu} - \hat{g}_{X_2} J_{B_3 - L_\tau}^\mu \hat{X}_{2\mu}
\end{eqnarray}
where $\hat{X}_1$ and $\hat{X}_2$ are the gauge bosons which belong to gauged $U(1)_{L_\mu - L_\tau}$ and $U(1)_{B_3 - L_\tau}$ in the gauge eigenbasis, respectively, and 
\begin{eqnarray}
J_{L_\mu - L_\tau}^\mu & = & \left ( \bar{\ell}_{2 L} \gamma^\mu \ell_{2 L} + \bar{\mu}_R \gamma^\mu \mu_R + \bar{N}_{2R} \gamma^\mu N_{2R} \right ) - \left ( \bar{\ell}_{3 L} \gamma^\mu \ell_{3 L} + \bar{\tau}_R \gamma^\mu \tau_R + \bar{N}_{3R} \gamma^\mu N_{3R} \right ), \ \ \ \ \ \ \ \ \ \\
J_{B_3 - L_\tau}^\mu & = & \frac{1}{3} \left ( \bar{q}_{3L} \gamma^\mu q_{3L} + \bar{t}_R \gamma^\mu t_R + \bar{b}_R \gamma^\mu b_R \right ) - \left ( \bar{\ell}_{3 L} \gamma^\mu \ell_{3 L} + \bar{\tau}_R \gamma^\mu \tau_R + \bar{N}_{3R} \gamma^\mu N_{3R} \right ) \ \ \ \ \ \
\end{eqnarray}
are the conserved currents of $U(1)_{L_\mu - L_\tau}$ and $U(1)_{B_3 - L_\tau}$, respectively.\footnote{$U(1)_{B_3 - L_\tau}$, flavoured $B-L$ for third generation fermions, has been considered to resolve the lepton universality violation in $R_K^{(*)}$ from the measurement of rare $B$ meson decays $B^0 \rightarrow K^{*0} l^- l^+$ ($l=e,\mu$) with TeV scale $X$ gauge boson and heavy vector-like fermions \cite{Alonso:2017uky}.} Here, $q_{3L} = (t \ \, b)_L^T$ is the third generation left-handed quark doublet in the SM.

Similar to single $X$ gauge boson cases, one can impose the general kinetic mixing between SM gauge bosons ($\gamma$ and $Z$) and new gauge bosons $\hat{X}_{1,2}$ with dimension-four operators $\epsilon_{\gamma X_i} \hat{B}_{\mu \nu} \hat{X}_i^{\mu \nu}$ $(i=1,2)$ although sizeable values of $\epsilon_{\gamma X_i}$ are constrained by dark photon searches \cite{Lees:2017lec, Banerjee:2017hhz}. See Appendix \ref{appendix_diagonalization} for the detailed formulation in the presence of generic kinetic and mass mixing. To obtain the physical spectrum and interactions, we diagonalize them from gauge eigenstates to mass eigenstates. As a result, we obtain a simple pair of light gauge bosons as
\begin{eqnarray}
\mathcal{L} & \supset & \mathcal{L}_{\rm SM} + \sum_{i=1,2} \left ( - \frac{1}{4} X_{i \mu \nu} X_i^{\mu \nu} + \frac{m_{X,i}^2}{2} X_{i \mu} X_i^\mu - g_{X,i} J_i^\mu X_{i\mu} \right )
\end{eqnarray}
where $J_i^\mu = J_{L_\mu - L_\tau}^\mu + \epsilon_i J_{B_3 - L_\tau}^\mu \ (i=1,2)$. Here, the ratios $\epsilon_i$ and each couplings $g_{X,i}$ are determined by the model parameters $\hat{M}_i^2$, $\delta \hat{M}_{12}^2$, $\hat{g}_{X_i}$ and $\epsilon_{12}$ for gauge eigenstates $\hat{X}_i$. In this work, we focus on the phenomenological setup of an effectively light gauge boson in the ranges of $100 \text{ MeV} \lsim m_X \lsim 165$ MeV with a gauge coupling $ g_{X}$ to $(L_\mu - L_\tau) + \epsilon (B_3 - L_\tau)$ current where $\epsilon$ is a small ratio between muon and top quark couplings, rather than the complete two gauge boson construction starting from the gauge eigenstates.

\subsection{Explanation of KOTO events}
\label{subsec:Width_Model2}

\begin{figure}[h]
\centering
\subfloat[$i \mathcal{M}_a^{B_3}$]
{\includegraphics[width=0.32\textwidth]{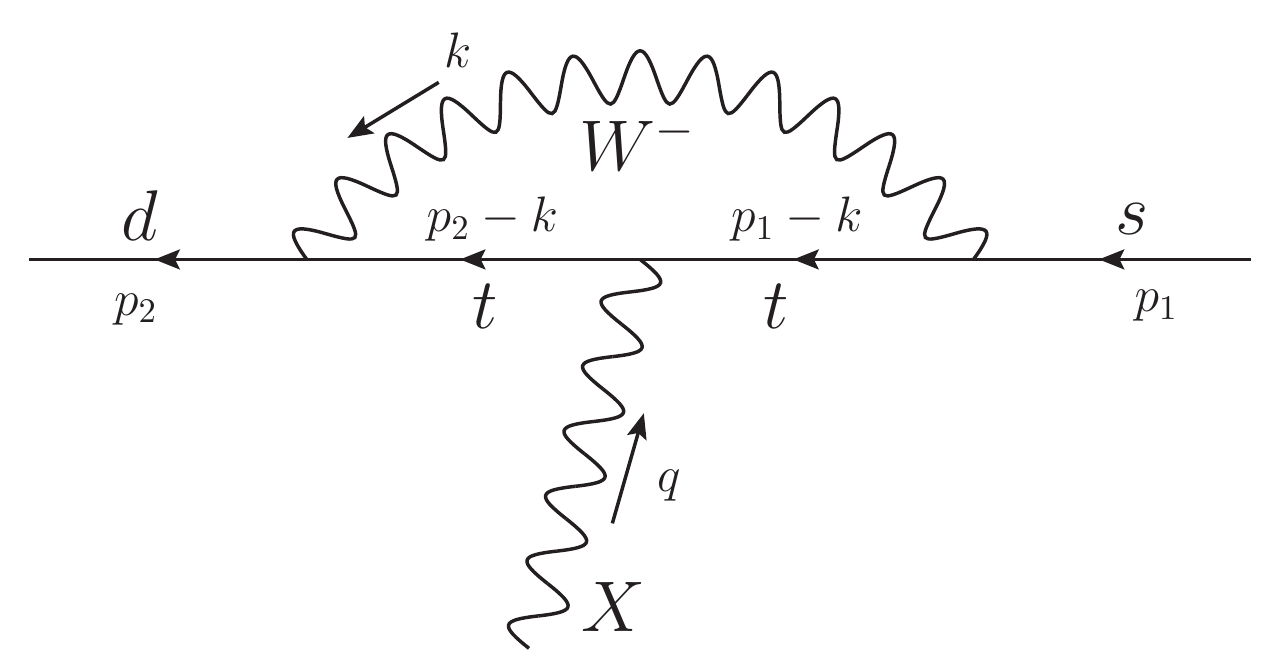}\label{fig_FCNC_B3_dsX_diagram_a}} \,
\subfloat[$i \mathcal{M}_b^{B_3}$]
{\includegraphics[width=0.32\textwidth]{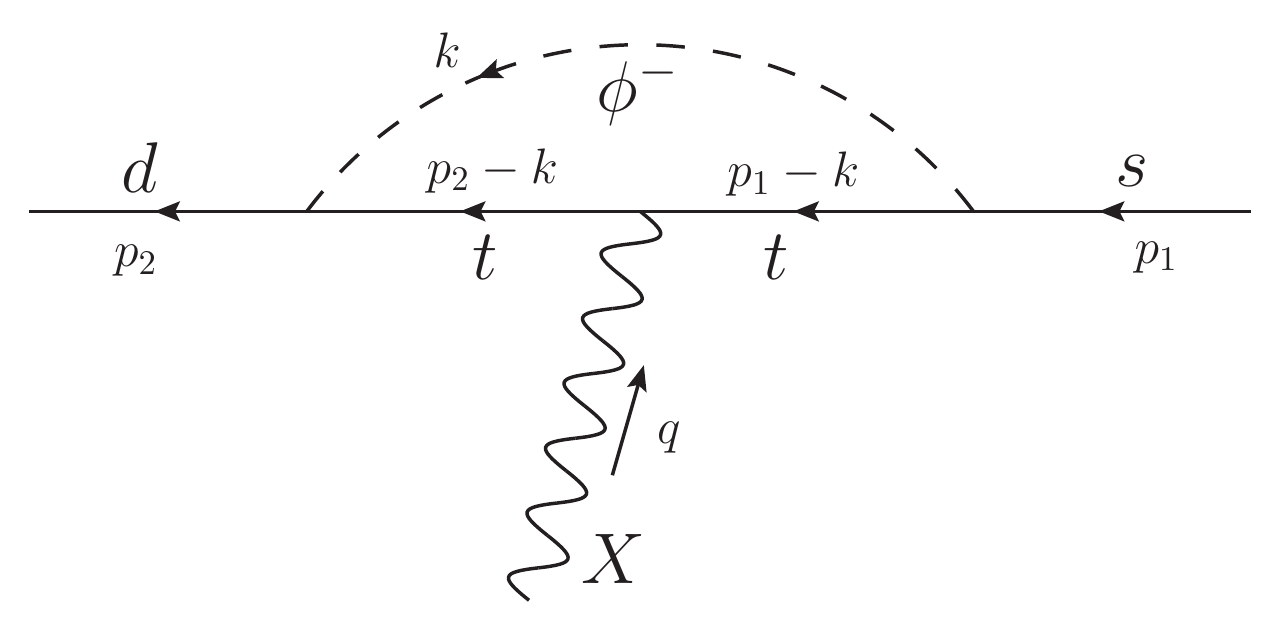}\label{fig_FCNC_B3_dsX_diagram_b}} \,
\subfloat[$\bar{u}_i \gamma^\mu P_L u_j X_\mu$]
{\includegraphics[width=0.32\textwidth]{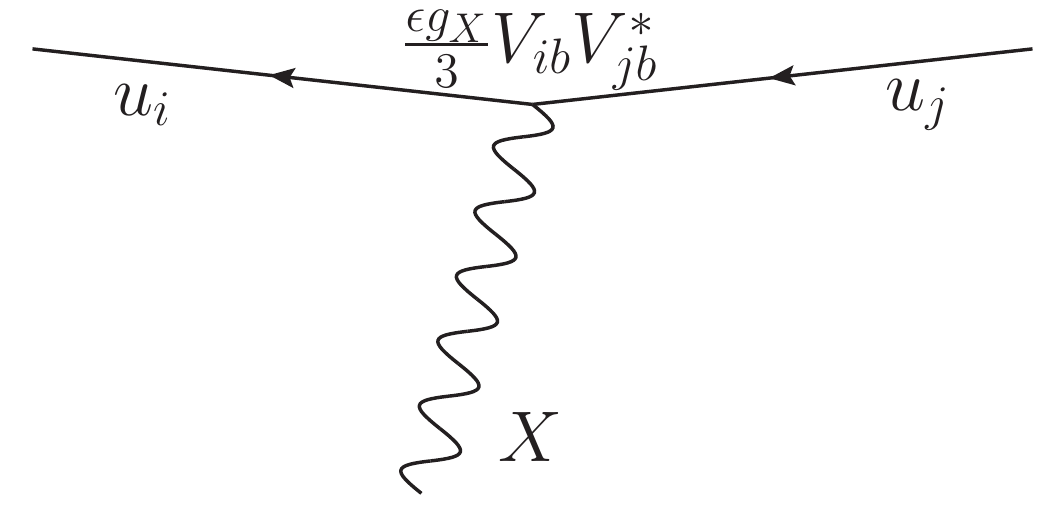}\label{fig_FCNC_B3_ucX_diagram_c}} \,
  \caption{The diagrams which contribute to the loop-induced $dsX$ FCNC process with $X$ gauge boson coupled to $B_3$ (model II). The loop-induced down-type FCNC (Left and Middle panels) and tree-level up-type FCNC (Right panel) are shown.} \label{fig_FCNC_B3_dsX_diagrams}
\end{figure}

In the presence of $X$ coupled to SM top quark, it significantly enhances FCNC at one-loop level. This is contrary to the SM photon case, because there is no cancellation 
among the diagrams from the symmetry. In the presence of $B_3$ coupling, effective FCNC couplings are
\begin{eqnarray}
\mathcal{L}_{\rm eff} & \supset & \frac{1}{3} \frac{g^2 \epsilon g_X V_{ts} V_{td}^*}{16\pi^2} \cdot F_1 (x_t) \cdot [\bar{d}_L \gamma^\mu s_L ] X_\mu + \frac{1}{3} \frac{g^2 \epsilon g_X V_{tb} V_{ts}^*}{16\pi^2} \cdot F_1 (x_t) \cdot [\bar{s}_L \gamma^\mu b_L ] X_\mu + \text{h.c.} \nonumber \\ 
& \equiv & g_{dsX}^{\rm eff} |_{B_3} [\bar{d}_L \gamma^\mu s_L ] X_\mu + g_{sbX}^{\rm eff} |_{B_3} [\bar{s}_L \gamma^\mu b_L ] X_\mu + \text{h.c.}
\end{eqnarray}
where
\begin{eqnarray}
F_1 (x_t) & \simeq & \frac{7x_t - 5 x_t^2 - 2 x_t^3 + x_t(x_t+2)^2 \ln x_t}{4(x_t-1)^2}
\end{eqnarray}
is the loop function of $X$ gauge boson induced penguin diagram in the limit $m_{d,s,b,X} \ll m_{W,t}$ with $x_t = m_t^2/m_W^2$ (See Appendix \ref{appendix_oneloop_FCNC}). We show the diagrams that contribute to down-type $s\rightarrow d X$ FCNC transition in Fig. \ref{fig_FCNC_B3_dsX_diagram_a} and Fig. \ref{fig_FCNC_B3_dsX_diagram_b}.

In terms of original $(L_\mu - L_\tau) + \epsilon (B_3 - L_\tau)$ gauge coupling $g_X$, FCNC couplings ($g_{sbX}^{\rm eff} |_{B_3}$ and $g_{dsX}^{\rm eff} |_{B_3}$) are given by
\begin{eqnarray}
g_{sbX}^{\rm eff} |_{B_3} & \simeq & (-2.73 \times 10^{-5} + 4.71 \times 10^{-7}i) \epsilon g_X, \label{downFCNC_B3_sb} \\
g_{dsX}^{\rm eff} |_{B_3} & \simeq & (-2.27 \times 10^{-7} - 8.86 \times 10^{-8}i) \epsilon g_X. \label{downFCNC_B3_ds}
\end{eqnarray}
up to $\mathcal{O}(\lambda^5)$ in the expansion of the Wolfenstein parameters. The upper bounds on $\epsilon g_X$ from Br($K^+ \rightarrow \pi^+ X$), Br($B^+ \rightarrow K^+ X$) and the required value for KOTO are
\begin{eqnarray}
\epsilon g_X & \simeq & 1.31 \times 10^{-5} \ \ \ \ \ \ \ \ \text{(KOTO desired FCNC coupling for $q^2 = m_{\pi^0}^2$)} \\
\epsilon g_X & \lsim & 5.16 \times 10^{-5} \ \ \ \ \ \ \ \ \text{(From Br($K^+ \rightarrow \pi^+ X$) upper limit for $q^2 = m_{\pi^0}^2$)} \\
\epsilon g_X & \lsim & 1.41 \times 10^{-4} \ \ \ \ \ \ \ \ \text{(From GN bound for $q^2 = m_{\pi^0}^2$)} \\
\epsilon g_X & \lsim & 4.05 \times 10^{-5} \ \ \ \ \ \ \ \ \text{(From Br($B^+ \rightarrow K^+ X$) upper limit for $q^2 = m_{\pi^0}^2$)}
\end{eqnarray}

Considering a $(L_\mu - L_\tau) + \epsilon (B_3 - L_\tau)$ gauge boson with $5 \times 10^{-4} \lsim g_X \lsim 10^{-3}$, $\epsilon \simeq 0.01 - 0.03$ and $m_X \simeq 100 - 165$ MeV, we have a simple interpretation for $(g-2)_\mu$ and KOTO events. We show this value of top quark coupling is consistent with other current constraints from other FCNC decays such as $K_L \rightarrow \mu^+ \mu^+$, $B_s \rightarrow \mu^+ \mu^+$ and neutral $K$, $B$, and $D$ meson mixings.

One can analogously consider $B_2$ (the baryon number of second generation) gauge coupling to make FCNC via charm quark contribution as $g_{dsX}^{\rm eff} |_{B_2} \sim \frac{1}{3} \frac{\epsilon g_X g^2 V_{cs} V_{cd}^*}{16\pi^2} F_1 (x_c)$, and obtain FCNC couplings
\begin{eqnarray}
g_{sbX}^{\rm eff} |_{B_2} & \simeq & -6.09 \times 10^{-8} \epsilon g_X, \\
g_{dsX}^{\rm eff} |_{B_2} & \simeq & (3.24 \times 10^{-7} - 1.93 \times 10^{-10} i) \epsilon g_X
\end{eqnarray}
up to $\mathcal{O}(\lambda^5)$ in the expansion of the Wolfenstein parameters again. However, it cannot provide a desired Br($K_L^0 \rightarrow \pi^0 X$) value, avoiding Br($K^+ \rightarrow \pi^+ X$) constraint at the same time
because
the imaginary part is three order of magnitude smaller than the real part in $g^{\rm eff}_{dsX}|_{B_2}$.
Similar to the minimal kinetic mixing case, shown in Section \ref{subsec:Kinetic_mixing}, charm quark contribution spoils loop-level FCNC explanation of KOTO excess without changing the mixing structure in the quark sector.

\subsection{Constraints}
\label{subsec:constraint_model2}

\begin{figure}[h]
\centering
\subfloat[$\epsilon = 0.012$]
{\includegraphics[width=0.49\textwidth]{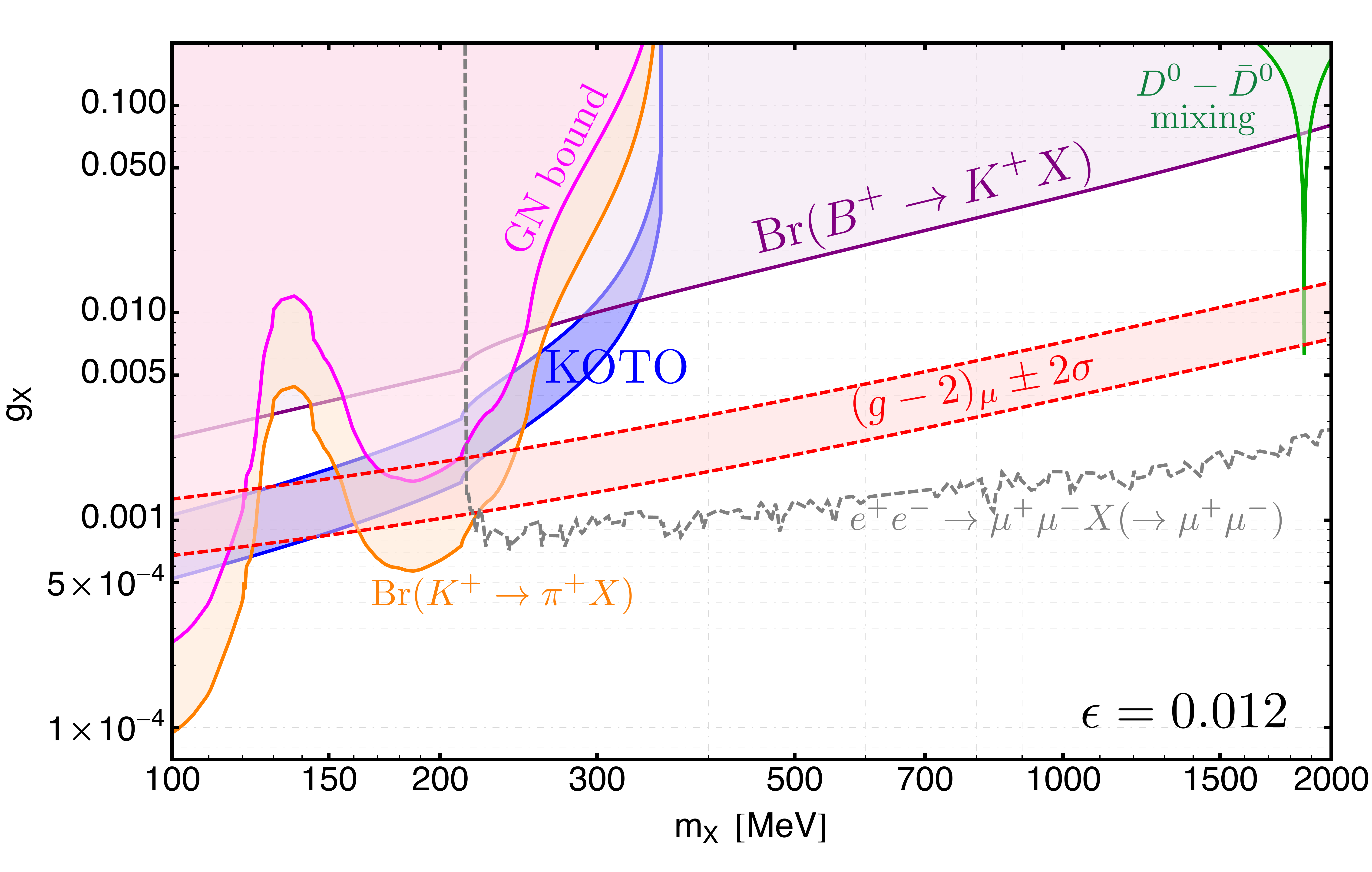}\label{fig_Model2param_01}} \,
\subfloat[$\epsilon = 0.035$]
{\includegraphics[width=0.49\textwidth]{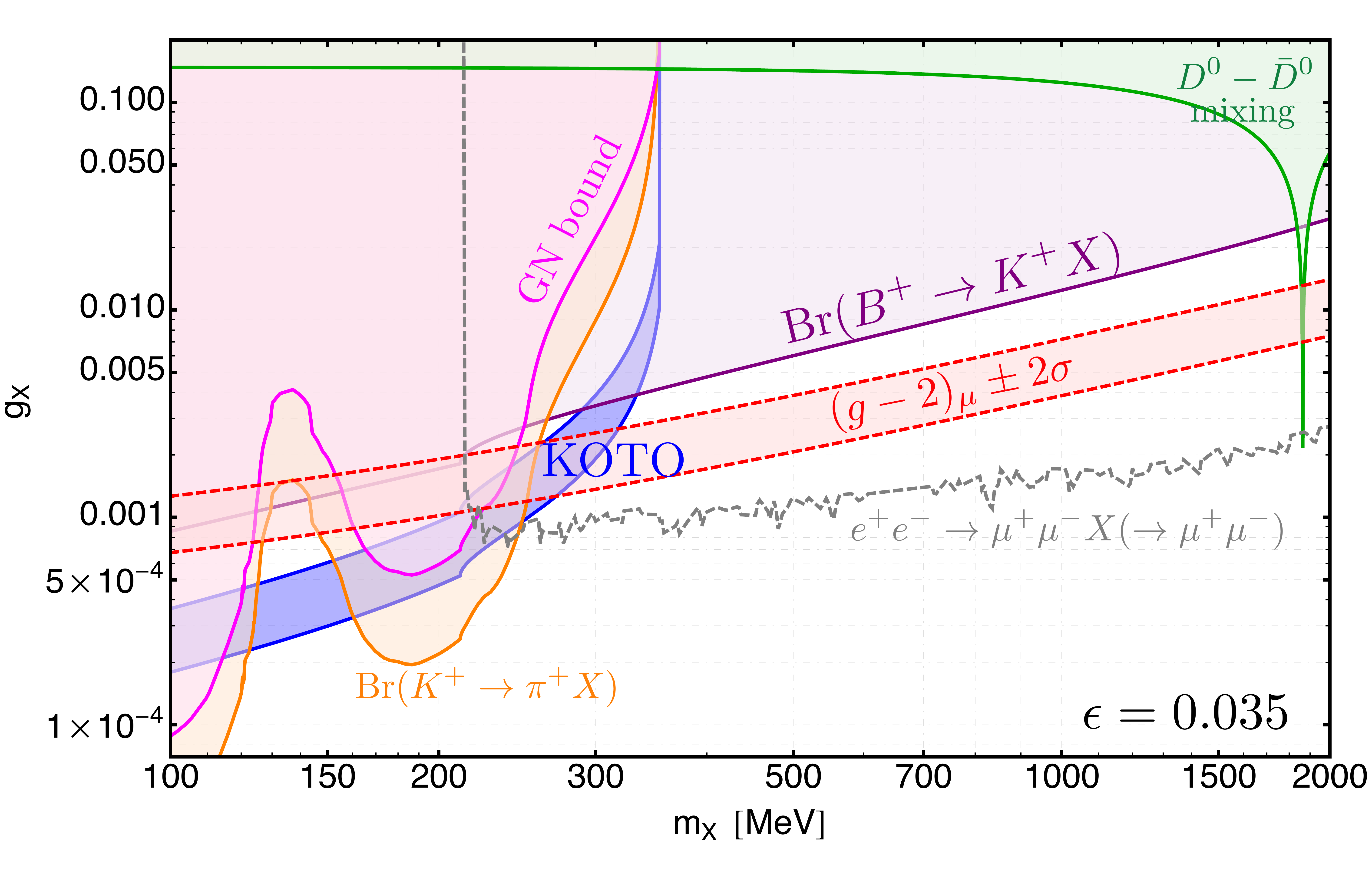}\label{fig_Model2param_02}}
  \caption{The preferred region for KOTO result and $(g-2)_\mu$ with a $(L_\mu - L_\tau) + \epsilon (B_3 - L_\tau)$ gauge boson $X$. All solid lines belong to $B_3$ coupling and dashed lines to $L_\mu$ coupling. Blue shaded band is the region for KOTO desired $\text{Br}(K_L^0 \rightarrow \pi^0 X)$ value. Red shaded band is the required value for $(g-2)_\mu$. The constraints from $\text{Br}(B^+ \rightarrow K^+ X)$ (purple), $\text{Br}(K^+ \rightarrow \pi^+ X)$ (orange), GN bound (magenta), $D^0 - \bar{D}^0$ mixing (green) and muonic force search in $4\mu$ channel (gray dashed) are also shown. We show two different cases of the ratio $\epsilon$ between the muonic and the hadronic couplings as $\epsilon = 0.012$ (Left panel) and $\epsilon = 0.035$ (Right panel). See the main text for details.} \label{fig_Model2param}
\end{figure}

In this section, we consider possible constraints
and summarize them in Fig. \ref{fig_Model2param}. 
In Fig. \ref{fig_Model2param}, we show the preferred region of parameters ($m_X$, $g_X$ and $\epsilon$) and the current experimental constraints. In model II, we have two allowed regions ($120 \text{ MeV} \lsim m_X \lsim 160 \text{ MeV}$ and $250 \text{ MeV} \lsim m_X \lsim 350 \text{ MeV}$) for the KOTO events, although higher mass region cannot explain $(g-2)_\mu$ simultaneously, due to the experimental constraints from $4\mu$ search from BaBar \cite{TheBABAR:2016rlg} and the search of the muonic force coupled to $b \rightarrow sX$ FCNC vertex at LHCb \cite{Aaij:2015tna}.

\subsubsection{$B^0 - \bar{B}^0$/$K^0 - \bar{K}^0$ and $D^0 - \bar{D}^0$ mixing}

In model II, we have loop-induced down-type FCNC couplings as Eq. (\ref{downFCNC_B3_sb}) and Eq. (\ref{downFCNC_B3_ds}), contributing to mixings of neutral mesons. Nevertheless, there are upper bounds from $B^0 - \bar{B}^0$ and $K^0 - \bar{K}^0$ mixings, which are converted into the Wilson coefficients of six-dimensional operators $(\bar{s}_L \gamma^\mu b_L)^2$ and $(\bar{s}_L \gamma^\mu d_L)^2$ respectively. The experimental upper bounds are \cite{Isidori:2010kg}
\begin{eqnarray}
| \text{Re} \ (\Lambda_{db}^{-2}) | & \ \leq \ & 3.3 \times 10^{-12} \text{ GeV}^{-2}, \\
| \text{Im} \ (\Lambda_{db}^{-2}) | & \ \leq \ & 1.0 \times 10^{-12} \text{ GeV}^{-2}, \\
| \Lambda_{sb}^{-2} | & \ \leq \ & 7.6 \times 10^{-11} \text{ GeV}^{-2}
\end{eqnarray}
for $B_s^0 - \bar{B}_s^0$/$B_d^0 - \bar{B}_d^0$ mixings as well as Eq. (\ref{KaonMixingRe}-\ref{KaonMixingIm}) for $K^0 - \bar{K}^0$ mixing. Due to the loop and CKM suppressions, they give very weak upper bounds, $\frac{\epsilon g_X}{3} \lsim 1.11$ from $B$ meson mixing and $\frac{\epsilon g_X}{3} \lsim 0.14$ from Kaon mixing, at $m_X \simeq m_{\pi^0}$.

In addition to down-type FCNC couplings ($g_{dsX}^{\rm eff} |_{B_3}$, $g_{sbX}^{\rm eff} |_{B_3}$ and $g_{dbX}^{\rm eff} |_{B_3}$) which are proportional to $\frac{g^2}{16\pi^2} V_{ti} V_{tj}^* F_1 (x_t)$, there are also tree-level up-type (left-handed) FCNC couplings due to $SU(2)_L$ gauge invariance as
\begin{eqnarray}
\mathcal{L}_{u_i u_j X} & \supset & \frac{\epsilon g_X}{3} \begin{pmatrix}
\bar{u} & \bar{c} & \bar{t} 
\end{pmatrix}_L \gamma^\mu
\begin{pmatrix}
0 & 0 & 0 \\
0 & 0 & 0 \\
0 & 0 & 1
\end{pmatrix}
\begin{pmatrix}
u \\
c \\
t
\end{pmatrix}_L X_\mu 
=
\frac{\epsilon g_X}{3} \begin{pmatrix}
\bar{u} & \bar{c} & \bar{t} 
\end{pmatrix}_L^m \mathcal{U}_L \gamma^\mu
\begin{pmatrix}
0 & 0 & 0 \\
0 & 0 & 0 \\
0 & 0 & 1
\end{pmatrix} \mathcal{U}_L^\dagger
\begin{pmatrix}
u \\
c \\
t
\end{pmatrix}_L^m X_\mu \nonumber \\
& = & \frac{\epsilon g_X}{3} \begin{pmatrix}
\bar{u} & \bar{c} & \bar{t} 
\end{pmatrix}_L^m \gamma^\mu
\begin{pmatrix}
| V_{ub} |^2 \ \ & V_{ub} V_{cb}^* \ \ & V_{ub} V_{tb}^* \\
V_{cb} V_{ub}^* \ \ & | V_{cb} |^2 \ \ & V_{cb} V_{tb}^* \\
V_{tb} V_{ub}^* \ \ & V_{tb} V_{cb}^* \ \ & | V_{tb} |^2
\end{pmatrix}
\begin{pmatrix}
u \\
c \\
t
\end{pmatrix}_L^m X_\mu
\end{eqnarray}
where $\mathcal{U}_L \mathcal{D}_L^\dagger = V_{\rm CKM}^{\rm SM}$ and we assume $\mathcal{U}_L = V_{\rm CKM}^{\rm SM}$, $\mathcal{D}_L = 1$ in our model. It generates sizeable tree-level up-type FCNC interactions
\begin{eqnarray}
g_{ucX}^{\rm eff} |_{B_3,\text{tree-level}} & \simeq & V_{ub} V_{cb}^*  \frac{\epsilon g_X}{3} \approx (1.62 \times 10^{-5} - 4.45 \times 10^{-5} i) \epsilon g_X, \\
g_{utX}^{\rm eff} |_{B_3,\text{tree-level}} & \simeq & V_{ub} V_{tb}^*  \frac{\epsilon g_X}{3} \approx (3.84 \times 10^{-4} - 1.06 \times 10^{-3} i) \epsilon g_X, \\
g_{ctX}^{\rm eff} |_{B_3,\text{tree-level}} & \simeq & V_{cb} V_{tb}^*  \frac{\epsilon g_X}{3} \approx 1.40 \times 10^{-2} \epsilon g_X.
\end{eqnarray}

The coupling $g_{ucX}^{\rm eff} |_{B_3,\text{tree-level}}$ can be constrained by $D^0 - \bar{D}^0$ mixing as \cite{Isidori:2010kg}
\begin{eqnarray}
\left | \text{Re} \left \{ \frac{(g_{ucX}^{\rm eff})^2}{m_{D^0}^2 - m_X^2} \right \} \right | & \lsim & 5.6 \times 10^{-13} \text{ GeV}^{-2}, \\
\left | \text{Im} \left \{ \frac{(g_{ucX}^{\rm eff})^2}{m_{D^0}^2 - m_X^2} \right \} \right | & \leq & 1.0 \times 10^{-13} \text{ GeV}^{-2} 
\end{eqnarray}
which are the real and imaginary part of the Wilson coefficient for the operator $(\bar{u}_L \gamma^\mu c_L)^2$. At $m_X \simeq m_{\pi^0}$, the constraints are translated into $\epsilon g_X \lsim 1.55 \times 10^{-2}$, which is not enough to constrain the KOTO required value.

The couplings $g_{utX}^{\rm eff} |_{B_3,\text{tree-level}}$, $g_{ctX}^{\rm eff} |_{B_3,\text{tree-level}}$ makes a FCNC decay of top quark. However, the branching ratio is much smaller than current experimental sensitivities from LHC searches.

\subsubsection{$\Gamma_{D^+}$ and $D^+ \rightarrow \pi^+ X$}

The coupling $g_{ucX}^{\rm eff} |_{B_3,\text{tree-level}}$ also promotes FCNC decay of the charged $D$ meson. The branching ratio of the decay $D^+ \rightarrow \pi^+ X$ is given by
\begin{eqnarray}
\text{Br}(D^+ \rightarrow \pi^+ X) & = & \frac{1}{\Gamma_{D^+,\text{total}}} \frac{1}{144\pi} \frac{m_{D^+}^3}{m_X^2} | F_+ (m_X^2) |^2 | g_{ucX}^{\rm eff} |^2
\end{eqnarray}
where $F_+ (q^2) = \frac{f_D}{f_\pi} \frac{g_{D^* D \pi}}{1 - q^2/m_{D^*}^2}$ is the form factor obtained from chiral perturbation theory of heavy hadrons \cite{Burdman:2003rs}. We use $f_D = 200$ MeV, $f_\pi = 130$ MeV and $g_{D^+ D \pi} = 0.59$ in our calculation, following the analysis given in Ref. \cite{Babu:2017olk}. We set our upper bound by requiring $\Gamma (D^+ \rightarrow \pi^+ X) < \Gamma_{D^+,\text{total}} - \Gamma_{D^+,K^0}$ using the inclusive value of the branching ratios, to avoid a significant modification of the total width of $D^+$ meson.\footnote{In our model, $X$ gauge boson dominantly decays into neutrino pair below the muon threshold. In this case, the invisible FCNC decay $D^+ \rightarrow \pi^+ X$ is suffered from the lack of $D^+$ reconstruction. Conservatively, we can impose the bound $\Gamma(D^+ \rightarrow \pi^+ X) < \Gamma_{D^+,\text{total}} - \Gamma_{D^+,K^0}$, since it cannot change the inclusive $K^0$ and $\bar{K}^0$ decay modes (Br($D^+ \rightarrow K^0, \bar{K}^0 + \text{anything}$)$\ \approx 61 \%$) significantly, which are not affected by the decay mode $D^+ \rightarrow \pi^+ X$, as pointed out in \cite{Babu:2017olk}.} At $m_X \simeq m_{\pi^0}$, it gives only a weak upper bound $\epsilon g_X \lsim 1.31 \times 10^{-2}$,
and thus not sensitive to KOTO and $(g-2)_\mu$ preferred region.

\subsubsection{$B_{d,s} \rightarrow \mu^+ \mu^-$/$K_L \rightarrow \mu^+ \mu^-$}

Before we go through following detailed analysis, 
we provide brief results of this subsection here.
For rare meson decays $K_L/B_s/B_d \rightarrow \mu^+ \mu^-$, 
the upper bound on FCNC couplings are about $\epsilon g^2_X \lesssim \mathcal{O}(10^{-5})$ 
and thus are insensitive to our bulk part parameter region, 
$\epsilon g^2_X \sim \mathcal{O}(10^{-8})$. 
Since the dominating uncertainties come from theoretical calculations, the upper bounds are determined by the condition where 
the $X$ boson contribution does not exceed the SM contribution for each decay channel.

For $K_L \rightarrow \mu^+ \mu^-$, as in Section \ref{subsubsec:KLtomumu_VLQ}, we write down the short-distance part of the effective hamiltonian as
\begin{eqnarray}
\mathcal{H}^{\rm SM}_{\rm eff}&=& -\frac{G_F}{\sqrt{2}}\frac{\alpha_{\rm EM}}{2\pi \sin^2 \theta_w}
\left( V^*_{cs}V_{cd}Y_{\rm NL}+V^*_{ts}V_{td}Y(x_t) \right)
[\bar{s}\gamma^{\mu}P_L d][\bar{\mu}\gamma_{\mu}P_L \mu]+{\rm h.c.}\,, \nonumber \\
\mathcal{H}^{B_3}_{\rm eff}&=& \left ( \frac{g^2 V_{td} V_{ts}^*}{16\pi^2}F_1 (x_t) \frac{\epsilon g_X^2}{3} \right ) \frac{1}{m^2_{K_L}-m^2_X}
[\bar{s}\gamma^{\mu}P_L d][\bar{\mu}\gamma_{\mu}P_L \mu]+{\rm h.c.}\,,
\end{eqnarray}
and the upper bound is given by demanding that the new physics contribution is smaller than the SM prediction value, as follows:
\begin{eqnarray}
\Bigl | \frac{\epsilon g_X^2}{3} \frac{g^2 V_{ts} V_{td}^*}{16\pi^2} F_1(x_t) \Bigl ( \frac{1}{m_{K_L}^2 - m_X^2} \Bigr ) \Bigr | & < & \left | \frac{G_F}{\sqrt{2}} \frac{\alpha_{\rm EM}}{2\pi \sin^2 \theta_W} \left( V^*_{cs}V_{cd}Y_{\rm NL}+V^*_{ts}V_{td}Y(x_t) \right) \right | \nonumber \\
\end{eqnarray}
and it gives $\epsilon g_X^2 \lsim 1.72 \times 10^{-5}$ for $m_X \simeq m_{\pi^0}$. 
The preferred values of $\epsilon \simeq 0.026$ and $g_X \simeq 5 \times 10^{-4}$ for KOTO and $(g-2)_\mu$ gives $\epsilon g^2_X \simeq \mathcal{O}(10^{-8})$,
therefore the $K_L \rightarrow \mu^+ \mu^-$ decay branching ratio is not sensitive to our model parameters.

For $B_s \rightarrow \mu^+ \mu^-$, we have
\begin{eqnarray}
\mathcal{H}^{\rm SM}_{\rm eff} & = & -\frac{G_F}{\sqrt{2}}\frac{\alpha_{\rm EM}}{2\pi \sin^2 \theta_w} V^*_{tb} V_{ts} Y(x_t) [\bar{s}\gamma^{\mu}P_L d][\bar{\mu}\gamma_{\mu}P_L \mu]+{\rm h.c.}\,, \nonumber \\
\mathcal{H}^{B_3}_{\rm eff} & = & \left ( \frac{g^2 V_{ts} V_{tb}^*}{16\pi^2}F_1 (x_t) \frac{\epsilon g_X^2}{3} \right ) \frac{1}{m^2_{B_s}-m^2_X}
[\bar{s}\gamma^{\mu}P_L d][\bar{\mu}\gamma_{\mu}P_L \mu]+{\rm h.c.}\,,
\end{eqnarray}
and
\begin{eqnarray}
\Bigl | \frac{\epsilon g_X^2}{3} \frac{g^2 V_{ts} V_{tb}^*}{16\pi^2} F_1(x_t) \Bigl ( \frac{1}{m_{B_s}^2 - m_X^2} \Bigr ) \Bigr | & < & \left | \frac{G_F}{\sqrt{2}} \frac{\alpha_{\rm EM}}{2\pi \sin^2 \theta_W} V^*_{tb} V_{ts} Y(x_t) \right |
\end{eqnarray}
with the same criterion. The upper bound is $\epsilon g_X^2 \lsim 1.92 \times 10^{-3}$ for $m_X \simeq m_{\pi^0}$, which is even weaker than Kaon constraints. For $B_d$ meson, the branching ratio is given by
\begin{eqnarray}
\frac{\text{Br}(B_d \rightarrow \mu^+ \mu^-)}{\text{Br}(B_s \rightarrow \mu^+ \mu^-)} & \simeq & \frac{\Gamma_{B_s,\text{total}}}{\Gamma_{B_d,\text{total}}} \frac{m_{B_d}}{m_{B_s}} \frac{F_{B_d}^2}{F_{B_s}^2} \frac{|V_{td}|^2}{|V_{ts}|^2}
\end{eqnarray}
where $F_{B_d} \simeq F_{B_s} \approx 210$ MeV. Thus, $B_d$ meson decay gives a similar upper limit value of the coupling $g_X$.

\subsubsection{Expected sensitivities in future experiments}

\begin{figure}[h!]
\centering
{\includegraphics[width=0.96\textwidth]{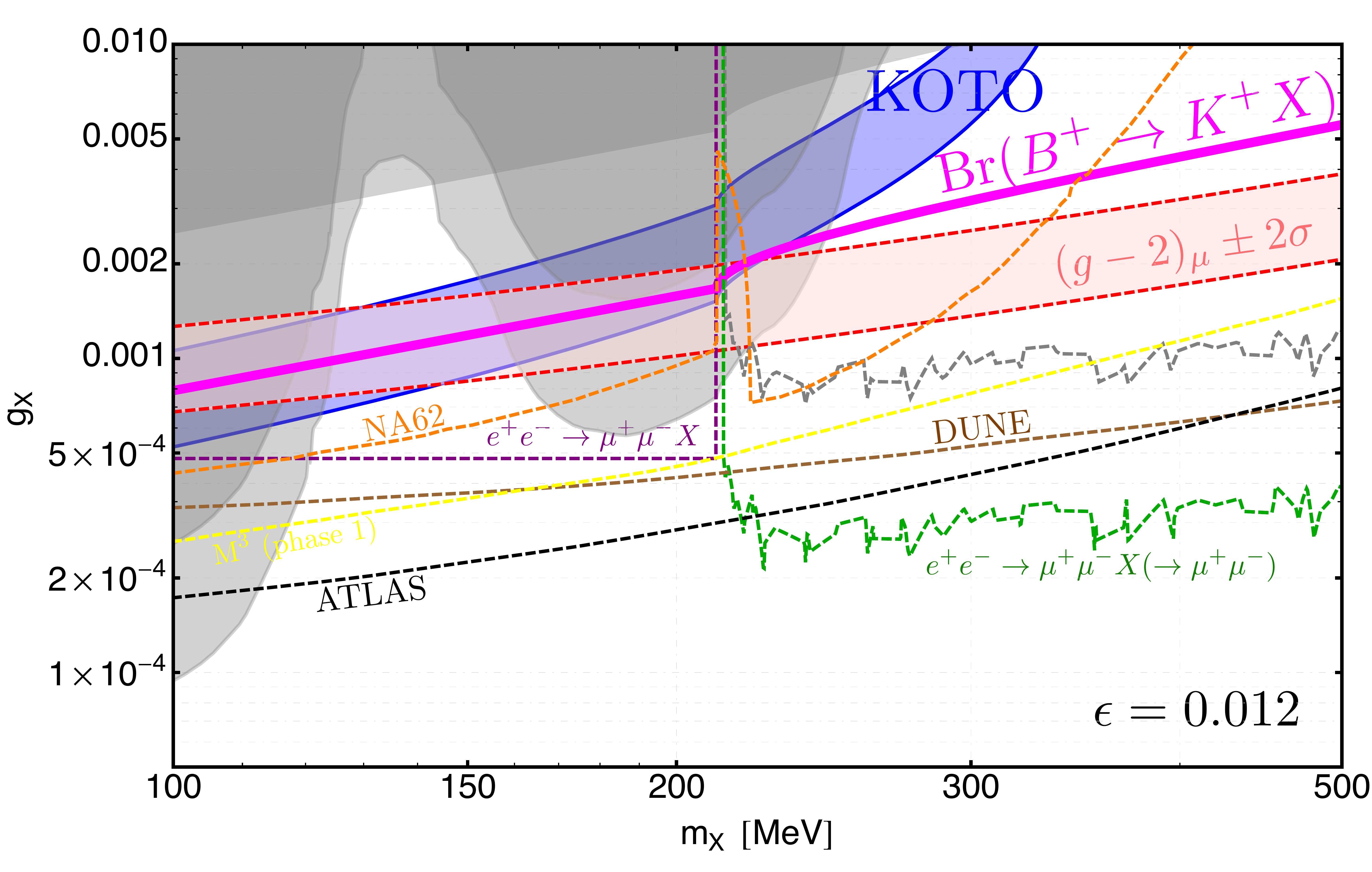}\label{fig_Model2param_FutureSensitivity_01}} \,
  \caption{The sensitivity limit expected in future experiments, for model II with a $(L_\mu - L_\tau) + \epsilon (B_3 - L_\tau)$ gauge boson $X$. All solid lines belong to $B_3$ coupling and dashed lines to $L_\mu$ coupling. Blue shaded band is the region for KOTO desired $\text{Br}(K_L^0 \rightarrow \pi^0 X)$ value. Red shaded band is the required value for $(g-2)_\mu$. From the existing upper limits of Belle and BaBar, we show the Belle II (with the data of 50 ab$^{-1}$ integrated luminosity) expected upper limits from i) $\text{Br}(B^+ \rightarrow K^+ X)$ (magenta) for hadronic coupling and muonic force searches using ii) $4\mu$ channel (green) and iii) $\mu^- \mu^+ X(\rightarrow \text{inv.})$ channel (purple) \cite{Jho:2019cxq, Graziani2019}. We also show the limit from for $\nu$-trident production at DUNE (brown) \cite{Ballett:2019xoj}, Kaon decays ($K^+ \rightarrow \mu \nu X$) at NA62 (orange) \cite{Krnjaic:2019rsv}, M${}^3$ (yellow) \cite{Kahn:2018cqs} and ATLAS (black) \cite{Galon:2019owl} for comparison. We set $\epsilon = 0.012$ as an example case.} \label{fig_Model2param_future_sensitivity}
\end{figure}

The most promising way to probe the KOTO preferred parameter region in model II is the rare decay of the charged $B$ meson ($B^+ \rightarrow K^+ X(\rightarrow \text{inv.})$) search at Belle II. The strongest upper bound on $\text{Br}(B^+ \rightarrow K^+ X)$ comes from Belle \cite{Chen:2007zk} and BaBar \cite{delAmoSanchez:2010bk, Lees:2013kla}, which corresponds to $\text{Br}(B^+ \rightarrow K^+ X) \lsim (1.3-1.6) \times 10^{-5}$ with the data of 492 ab$^{-1}$ and 418 ab$^{-1}$, respectively. By a simple rescaling for the upper limit as $g_X^{\rm upper.} \propto (\int dt \mathcal{L})^{-1/4}$, we show the expected limit at Belle II in Fig.~\ref{fig_Model2param_future_sensitivity}. We also include the expected limits on the muonic force from Belle II using $4\mu$ channel \cite{TheBABAR:2016rlg} and $\mu^- \mu^+ X(\rightarrow \text{inv.})$ channel \cite{Jho:2019cxq, Graziani2019}, and neutrino-trident production at DUNE \cite{Ballett:2019xoj} for a $(L_\mu - L_\tau) + \epsilon (B_3 - L_\tau)$ gauge boson $X$, 
Kaon decays ($K^+ \rightarrow \mu \nu X$) at NA62 \cite{Krnjaic:2019rsv}, M${}^3$ (Muon Missing Momentum) based at Fermilab \cite{Kahn:2018cqs}, and ATLAS detector as muon fixed-target experiment \cite{Galon:2019owl} for comparison.
We show the expected sensitivities in Fig.~\ref{fig_Model2param_future_sensitivity}.

For both muonic ($L_\mu$) and hadronic ($B_3$) coupling, most of $(g-2)_\mu$ and KOTO desired region can be probed by Belle II through $\text{Br}(B^+ \rightarrow K^+ X)$ and $4\mu$ channel searches, with the data of 50 ab$^{-1}$ integrated luminosity. Note that we assume similar systematic uncertainties in $\text{Br}(B^+ \rightarrow K^+ X)$ and $4\mu$ channel search of muonic force. Thus, the actual limit could be different from our estimation, depending on experimental environment at future experiments.

\section{Summary and Conclusion}
\label{sec:summary_conclusion}

The long-standing $(g-2)_\mu$ anomaly and recent J-PARC KOTO event excess can be explained in single framework by a light ($m_X < 2 m_\mu$) gauge boson $X$, 
where its mass is near the neutral pion mass in order to avoid the 
stringent GN bound and ${\rm Br}(K^+ \to \pi^+ +\text{invisible})$ upper limit. The $X$ boson has to couple to both lepton and quark sectors, and we investigated 
possibilities from two model frameworks, $i)$ gauged ${L_\mu - L_\tau}$ with heavy VLQs, $ii)$ gauged ${L_\mu - L_\tau}+\epsilon (B_3-L_\tau)$ with mixing of two gauge bosons.
Both frameworks provide allowed parameter regions for $(g-2)_\mu$ and KOTO, and satisfy the current experimental constraints. We would like to summarize our results in the following list.

\begin{itemize}
\item The simple model from $U(1)_X$ boson mixing with SM photon  
cannot interpret the KOTO event, meanwhile satisfying the constraint 
from ${\rm Br}(K^+ \to \pi^+ +\text{invsible})$.
\item In gauged ${L_\mu - L_\tau}$ with heavy VLQs, 
the $(g-2)_\mu$ prefers gauge coupling $g_X=5\times 10^{-4}$, 
and KOTO event excess requires 2 TeV mass VLQs carrying 
complex FCNC Yukawa couplings of 
${\rm Im}(Y_{Qs}Y^*_{Qd})\simeq 2.74\times 10^{-7}$,
which is compatible with constraints from $K^0-\bar{K}^0$ mixing, $K_L \to \mu^+\mu^-$,
and CKM unitarity.
\item The ${L_\mu - L_\tau}+\epsilon (B_3-L_\tau)$ gauge boson 
with $5 \times 10^{-4} \lsim g_X \lsim 10^{-3}$, 
$\epsilon \simeq 0.01 - 0.03$ and $m_X \simeq 120 - 160$ MeV provides 
simple interpretation for both $(g-2)_\mu$ and KOTO events.
Meanwhile, it satisfies the GN bound, ${\rm Br}(K^+ \to \pi^+ + \text{invisible})$, 
and ${\rm Br}(B^+ \to K^+ + \text{invisible})$ upper limits.
In near future, this preferred parameter region 
will be explored by the $B^+ \to K^+ + \text{invisible}$ search at Belle II.
On the other hand, the muonic force region will be tested by the 
$e^+ e^- \to \mu^+ \mu^- X \to \mu^+ \mu^- + \text{invisible}$ channel at Belle II, $\nu$-trident production at DUNE, Kaon decay at NA62, Muon beam dump experiment and muonic decay of $W$/$Z$ at ATLAS.
\item Another parameter region, 
the ${L_\mu - L_\tau}+\epsilon (B_3-L_\tau)$ gauge boson with 
$1 \times 10^{-3} \lsim g_X \lsim 3\times 10^{-3}$, 
$\epsilon \simeq 0.03-0.04$ and $m_X \simeq 250 - 350$ MeV 
interprets both $(g-2)_\mu$ and KOTO events.
But it has been excluded by $4\mu$ channel searches from BaBar,
since $m_X> 2m_\mu$ and thus muon decay channel is allowed. 
\end{itemize} 

The observation of $K_L^0 \rightarrow \pi^0 + (\text{invisible})$ decay events are based on the analysis of the 2016-2018 KOTO data, where the current sensitivity reaches a single event for $K_L$ branching ratio of $\sim \mathcal{O}(10^{-10})$. 
The enhanced data collected by KOTO experiment in 2019 is expected to improve the statistical uncertainty in near future \cite{Shinohara:2019}. Furthermore, several upcoming experiments on rare Kaon decays, such as KOTO step-2 \cite{Togawa:2013qxa} and KLEVER using CERN SPS beam for the $K_L$ production during the period of LHC Run 4~\cite{Ambrosino:2019qvz, Moulson:2019ifj}, have been proposed and the projected sensitivity can reach branching ratio of $\sim \mathcal{O}(10^{-13})$ so that it will fully cover the SM prediction $\sim \mathcal{O}(10^{-11})$. 
Combining with the various and extensive searches on the muonic force \cite{Kaneta:2016uyt,Araki:2017wyg,Jho:2019cxq,Krnjaic:2019rsv,Ballett:2019xoj,Kahn:2018cqs}, they will provide new probes of the models suggested in this work.

\acknowledgments
This work was supported by the National Research Foundation of Korea (NRF) grant funded by the Korean government (MSIP) (NRF-2018R1A4A1025334) and (NRF-2019R1A2C1089334). The work of SML was supported in part by the Hyundai Motor Chung Mong-Koo Foundation. The work of YSJ was supported by IBS under the project code, IBS-R018-D1.

\appendix

\section*{Appendix}

\section{Diagonalization of two hidden gauge bosons with generic kinetic and mass mixings}
\label{appendix_diagonalization}

The diagonalization method in the presence of two additional $U(1)$ gauge bosons with kinetic/mass mixings has been discussed in the Ref. \cite{Heeck:2011md}. In this section, we will present an analytic method of diagonalization without approximations and discuss the origin of $\epsilon$ factor in Model II. 

The neutral sector of the most general Lagrangian for ${G_{\rm SM}\times U(1)_{1} \times U(1)_{2}}$ spontaneously broken to ${SU(3)_c \times U(1)_{\rm EM}}$ after (several) higgsing is conveniently written as $\mathcal{L}=\mathcal{L}_{\rm kin} + \mathcal{L}_{\rm mass} +\mathcal{L}_{\rm mix}$ with
\begin{eqnarray}
\mathcal{L}_{\rm kin} & = & -\frac{1}{4} \left( \hat{W}_{\mu\nu}^3 \hat{W}^{3\mu\nu} + \sum_{i=0}^{2} \hat{K}_{ij} \hat{X}_{i \mu\nu}\hat{X}_i^{\mu\nu}   \right), \\
\mathcal{L}_{\rm mass} & = & \frac{1}{2} \left( \hat{M}_Z^2 \hat{Z}_\mu \hat{Z}^\mu + \sum_{i=1}^2 \hat{M}_{X_i}^2 \hat{X}_{i\mu} \hat{X}_i^\mu \right), \\
\mathcal{L}_{\rm mix} & = & m_1^2 \hat{Z}_\mu \hat{X}_1^\mu + m_2^2 \hat{Z}_\mu \hat{X}_2^\mu + m_3^2 \hat{X}_{1\mu} \hat{X}_2^\mu
\end{eqnarray}
where $\hat{K}_{ij} \equiv \delta_{ij} +\theta_{ij} $ with the kinetic mixing (off-diagonal) parameters $\theta_{ij} (=\theta_{ji}) \equiv \sin \theta_k |\epsilon_{kij}|$ where the Levi-Civita symbol is defined with $i,j$ and $k$ runs from $0$ to $2$ and $\epsilon_{012}=1$. We denote $\hat{X}_{0} \equiv \hat{B}$, $\hat{Z} = \hat{c}_W \hat{W}_3 -\hat{s}_W \hat{B}$ and $\hat{A}=\hat{c}_W \hat{B} +\hat{s}_W \hat{W}_3$ where $\hat{W}_3$ denote the third component of the ${SU(2)}$ gauge fields and $\hat{c}_W =\cos \theta_W$ and $\hat{s}_W=\sin\theta_W$ are cosine and sine of the Weinberg angle. We conveniently write $\theta_0=\alpha$, $\theta_1=\beta$ and $\theta_2=\gamma$ and $s_\eta=\sin \eta$, $c_\eta=\cos\eta$ and  $t_\eta=\tan \eta$ for $\eta=\alpha, \beta,\gamma$ below. The mass and mixing terms are collectively written as
\begin{eqnarray}
\mathcal{L}_{\rm mass+mix} & = & \begin{pmatrix} 
\hat{A}_\mu & \hat{Z}_\mu & \hat{X}_{1 \mu} & \hat{X}_{2\mu}
\end{pmatrix} 
\cdot \hat{\cal M}^2 \cdot 
\begin{pmatrix} 
\hat{A}^\mu \\
\hat{Z}^\mu \\
\hat{X}_1^\mu \\
\hat{X}_2^\mu
\end{pmatrix},
\end{eqnarray}
where the mass matrix
\begin{eqnarray}
\hat{\mathcal{M}}^2 =
\begin{pmatrix}
0 & 0 & 0 & 0 \\ 
0 & \hat{M}_Z^2 & m_1^2& m_2^2 \\ 
0 & m_1^2 & \hat{M}_{X_1}^2 & m_3^2 \\ 
0 & m_2^2 & m_3^2 & \hat{M}_{X_2}^2
\end{pmatrix}.
\end{eqnarray}

\subsection{$ 2 \times 2 $ Case}

As one of the most simplest case, let us set $m_1=m_2=0$ by assuming decoupled Higgs processes and $m_3^2 =\delta \hat{M}^2$ denoting a mass mixing between two extra gauge bosons. We also consider kinetic mixing parameters to be $\alpha=\beta=0$, motivated by the fact that $ m_{1}, m_{2},\alpha,\beta $ are constrained by various experiments. Then it is straightforward to see that the problem reduces to a $2 \times 2$ matrix problem as the mass matrix becomes
\begin{eqnarray}
\hat{\mathcal{M}}^2 & = & \begin{pmatrix}
0 & 0 & 0 & 0 \\ 
0 & \hat{M}_Z^2 & 0 & 0 \\ 
0 & 0 & \hat{M}_{1}^2 & \delta \hat{M}^2 \\ 
0 & 0 & \delta \hat{M}^2 & \hat{M}_{2}^2
\end{pmatrix}.
\end{eqnarray}
To eliminate kinetic mixing, we redefine the fields
\begin{eqnarray}
\begin{pmatrix}
\hat{X}_1 \\ 
\hat{X}_2 
\end{pmatrix} & = & \begin{pmatrix}
1 & -t_\gamma \\ 
0 & 1/c_\gamma
\end{pmatrix}
\begin{pmatrix}
\tilde{X}_1 \\ 
\tilde{X}_2 
\end{pmatrix}
\end{eqnarray}
with transformed mass matrix
\begin{eqnarray}
\mu^{2} & = & 
\begin{pmatrix}
\hat{M}_1^2 & \delta \hat{M}^2/c_{\gamma} - \hat{M}_{1}^2 t_{\gamma} \\
\delta \hat{M}^{2}/c_{\gamma}-\hat{M}_{1}^{2}t_{\gamma}&\left[\hat{M}_{2}^{2}/c_{\gamma} +\left(	\hat{M}_{1}^{2} s_{\gamma} - 2 \delta \hat{M}^{2}	\right) t_{\gamma} \right]/c_{\gamma} 
\end{pmatrix}.
\end{eqnarray}

Two physical masses are given by the eigenvalues of the matrix $ \mu^{2} $ as
\begin{eqnarray}
M_{1}^{2} & = & \frac{1}{2c_{x}^{2}} \left(
\hat{M}_{1}^{2} + \hat{M}_{2}^{2} - 2 \delta \hat{M}^{2} s_{x}
\right) - \sqrt{  \left(
	\hat{M}_{1}^{2} + \hat{M}_{2}^{2} - 2 \delta \hat{M}^{2} s_{x}
	\right)^{2} + 4c_{x}^{2} \left(\delta \hat{M}^{4} - 
	\hat{M}_{1}^{2}\hat{M}_{2}^{2} \right)   } \nonumber \\
	& & \\
M_{2}^{2} & = & \frac{1}{2c_{x}^{2}} \left(
\hat{M}_{1}^{2} + \hat{M}_{2}^{2} - 2 \delta \hat{M}^{2} s_{x}
\right) + \sqrt{  \left(
	\hat{M}_{1}^{2} + \hat{M}_{2}^{2} - 2 \delta \hat{M}^{2} s_{x}
	\right)^{2} + 4c_{x}^{2} \left(\delta \hat{M}^{4} - 
	\hat{M}_{1}^{2}\hat{M}_{2}^{2} \right)   }.\nonumber \\
	& &
\end{eqnarray}
Corresponding orthogonal matrix made by eigenvectors is
\begin{eqnarray}
\mathcal{O}_{2 \times 2} & = & \begin{pmatrix}
\cos \phi & \sin \phi \\
-\sin \phi & \cos \phi
\end{pmatrix} 
\end{eqnarray}
where
\begin{eqnarray}
\tan 2\phi & = & \frac{2c_{\gamma} \left(\delta \hat{M}^{2} - s_{\gamma} \hat{M}_{1}^{2}\right)}{\hat{M}_{2}^{2} - c_{2\gamma} \hat{M}_{1}^{2} - 2 s_{\gamma} \delta \hat{M}^{2} }
\end{eqnarray}.
Therefore, the canonical fields $ (X_{1},X_{2})^{T} $ with no kinetic/mass mixings are
\begin{eqnarray}
\begin{pmatrix}
X_{1} \\ X_{2}
\end{pmatrix}
& = & \mathcal{O}_{2\times 2} \cdot 
\begin{pmatrix}
1 & - t_{\gamma} \\
0 & 1/c_{\gamma}
\end{pmatrix}
\begin{pmatrix}
\hat{X}_{1} \\ \hat{X}_{2}
\end{pmatrix}
\end{eqnarray}
or explicitly,
\begin{eqnarray}
X_{1} & = & c_{\phi} \hat{X}_{1} + \left(\frac{s_{\phi}}{c_{\gamma}} - c_{\phi} t_{x}\right) \hat{X}_{2} \\
X_{2} & = & -s_{\phi} \hat{X}_{1} + \left(\frac{c_{\phi}}{c_{\gamma}} + s_{\phi} t_{\gamma}\right) \hat{X}_{2}.
\end{eqnarray}
Inversely, we obtain
\begin{eqnarray}
\hat{X}_{1} & = & \left(	c_{\phi} + s_{\gamma}s_{\phi} \right) X_{1}+ \left( c_{\phi} s_{\gamma} - s_{\phi}\right) X_{2} \\
\hat{X}_{2} & = & c_{\gamma} s_{\phi} X_{1} + c_{\gamma} c_{\phi} X_{2}.
\end{eqnarray}

Taking the relation between the interaction eigenstates and the mass eigenstates into account, the interaction terms in Lagrangian using the mass eigenstates are given as follows.
\begin{eqnarray}
& & \hat{g}_{1} J_{L_{\mu} - L_{\tau}}\hat{X}_{1} + \hat{g}_{2} J_{B_{3} - L_{\tau}} \hat{X}_{2} \nonumber \\
& & = \hat{g}_{1} J_{L_{\mu} - L_{\tau}} \left( \left(	c_{\phi} + s_{\gamma}s_{\phi} \right) X_{1}+ \left( c_{\phi} s_{\gamma} - s_{\phi}\right) X_{2}	\right) + \hat{g}_{2} J_{B_{3} - L_{\tau}}  \left(  c_{\gamma} s_{\phi} X_{1} + c_{\gamma} c_{\phi} X_{2}	\right)
\end{eqnarray}
In this case, the $\epsilon$ factor in Model II (Section \ref{subsec:Model2_LmuLtau_B3Ltau_w_RHnu}) for $X_{1}$ is
\begin{eqnarray}
\epsilon & = & \frac{\hat{g_{2}}c_{\gamma}s_{\phi}}{\hat{g_{1}}\left(	c_{\phi} + s_{\gamma} s_{\phi}\right)}
\end{eqnarray}
as an example.
\subsection{$3 \times 3$ Case}
We will consider the most general $3 \times 3$ case without assuming the smallness of parameters. We first diagonalize the kinetic term by changing the basis $(\hat{B}, \hat{X}_1, \hat{X}_2)$ to $(B,X_1,X_2)$ as
\begin{eqnarray}
\begin{pmatrix}
\hat{B} \\ 
\hat{X}_1\\ 
\hat{X}_2 
\end{pmatrix} & = & \begin{pmatrix}
1 & -t_\alpha & k/D \\ 
0 & 1/c_\alpha & q/D \\ 
0 & 0 & c_\alpha/ D
\end{pmatrix}
\begin{pmatrix}
B\\ 
X_1\\ 
X_2
\end{pmatrix}
\end{eqnarray}
where $k =(t_\alpha s_\gamma -s_\beta/c_\alpha), q= (t_\alpha s_\beta -s_\gamma/c_\alpha)$ and $D=\sqrt{1-s_\alpha^2-s_\beta^2-s_\gamma^2+2s_\alpha s_\beta s_\gamma}$. With new basis, the kinetic term becomes
\begin{eqnarray}
\mathcal{L}_{\rm kin} 
& = & -\frac{1}{4}(W_{3\mu\nu}, B_{\mu\nu},X_{1\mu\nu},X_{2\mu\nu}) \cdot 1_{4\times 4} \cdot
\begin{pmatrix} 
W_3^{\mu\nu} \\
B^{\mu\nu} \\
X_1^{\mu\nu} \\
X_2^{\mu\nu}
\end{pmatrix} \\
& = & -\frac{1}{4} \begin{pmatrix} 
A_{\mu\nu} & Z_{\mu\nu} & X_{1\mu\nu} & X_{2\mu\nu} \end{pmatrix} \cdot 1_{4\times 4} \cdot
\begin{pmatrix} 
A^{\mu\nu} \\
Z^{\mu\nu} \\
X_1^{\mu\nu} \\
X_2^{\mu\nu}
\end{pmatrix}
\label{eq:kin}
\end{eqnarray}
where $A = \hat{c}_W B + \hat{s}_W W_3$ is the massless photon and 
$Z= \hat{c}_W W_3 -\hat{s}_W B$ is a massive boson. The parameters from physical Weinberg angle $ s_{W},~c_{W}  $ are connected by
\begin{eqnarray}
s_{W}c_{W} M_{1} = \hat{s}_{W}\hat{c}_{W} \hat{M}_{Z}.
\end{eqnarray}

Now we determine the mass eigenstates by diagonalizing $\mathcal{L}_{\rm mass}+\mathcal{L}_{\rm mix}$ 
\begin{eqnarray}
\mathcal{L}_{\rm mass+mix}
& = & \frac{1}{2} \begin{pmatrix} A_\mu & Z_\mu & X_{1\mu} & X_{2\mu} \end{pmatrix} \cdot \mathcal{M}^2 \cdot \begin{pmatrix} 
A^\mu \\
Z^\mu \\
{X}_1^\mu \\
{X}_2^\mu
\end{pmatrix} \\
& = & \frac{1}{2} \begin{pmatrix} A_\mu & Z_{1\mu} & Z_{2\mu} & Z_{3\mu} \end{pmatrix} \cdot \mathcal{M}_{\rm diag}^2 \cdot \begin{pmatrix} 
A^\mu \\
Z_1^\mu \\
{Z}_2^\mu \\
{Z}_3^\mu
\end{pmatrix}
\end{eqnarray}
where the unprimed symmetric mass matrix is obtained by the field redefinition
\begin{eqnarray}
\hat{Z} & = & Z+\hat{s}_W t_\alpha X_1-\hat{s}_W (k/D)X_2, \\ 
\hat{X}_1 & = & 1/c_\alpha X_1 + (q/D)X_2, \\
\hat{X}_2 & = & (c_\alpha/D) X_2
\end{eqnarray} 
from Eq.~(\ref{eq:kin}):
\begin{eqnarray}
{\cal M}^2 = \begin{pmatrix}
0 & 0 & 0 & 0 \\ 
0 & & & \\
0 & & \mu^{2} & \\
0 & & &
\end{pmatrix}
\end{eqnarray}
where the $3\times 3$ symmetric sub-matrix is
\begin{eqnarray}
\mu^{2} & = &
\begin{pmatrix}
\hat{M}_Z^2 &m_{1}^{2}/c_{\alpha} + \hat{M}_Z^2 \hat{s}_W t_\alpha  & \mu_{13}^2 \\ 
\hat{M}_Z^2 \hat{s}_W t_\alpha  & \hat{M}_{X_1}^2/c_\alpha^2 + \hat{s}_W t_\alpha (2m_{1}^{2} +\hat{M}_Z^2 \hat{s}_W s_{\alpha} )/c_{\alpha} & \mu_{23}^2 \\ 
\mu_{13}^2 & \mu_{23}^2 & \mu_{33}^2
\end{pmatrix}
\end{eqnarray}
with the parameters 
\begin{eqnarray}
\mu_{13}^2 & = & \left( \hat{M}_Z^2 \hat{s}_W (s_\beta-s_\alpha s_\gamma) +m_{1}^{2} (s_{\alpha}s_{\beta}-s_{\gamma}) +m_{2}^{2} c_{\alpha}^{2}\right) /(c_\alpha D),\\
\mu_{23}^2 & = & \Bigl ( \hat{M}_{X_1}^2 (s_\alpha s_\beta -s_\gamma)+\hat{M}_Z^2 \hat{s}_W^2 s_\alpha(s_\beta-s_\alpha s_\gamma) \nonumber \\ 
& & +m_{1}^{2} \hat{s}_{W} (s_{\beta} - 2 s_{\alpha} s_{\gamma} + s_{\beta} s_{\alpha}^{2})
+ m_{2}^{2} \hat{s}_{W} s_{\alpha}c_{\alpha}^{2} + m_{3}^{2} c_{\alpha}^{2} \Bigr )/(c_\alpha^2 D),\\
\mu_{33}^2 & = & \Bigl ( \hat{M}_{X_2}^2 c_\alpha^4 + \hat{M}_{X_1}^2 (s_\gamma -s_\alpha s_\beta)^2 + \hat{M}_{Z}^2 \hat{s}_W^2 (s_\beta-s_\alpha s_\gamma)^2
-2m_{1}^{2} \hat{s}_{W}(s_{\alpha}s_{\beta} - s_{\gamma})(s_{\alpha}s_{\gamma}-s_{\beta}) \nonumber \\
& & + 2m_{2}^{2} c_{\alpha}^{2} \hat{s}_{W}(s_{\beta} - s_{\alpha}s_{\gamma})
+2m_{3}^{2} c_{\alpha}^{2}(s_{\alpha}s_{\beta}-s_{\gamma}) \Bigr )/(c_\alpha^2 D^2).
\end{eqnarray}

Because the matrix is symmetric and real, it can be diagonalized by an orthogonal matrix ${\cal O}$ as ${\cal O}^T {\cal M}^2 {\cal O} ={\cal M}^2_{\rm diag}$. In particular, the photon remains massless, the orthogonal matrix has the form:
\begin{eqnarray}
\mathcal{O} = \begin{pmatrix}
1 & 0 & 0 & 0 \\
0 & & & \\
0 & & {\cal O}_{3\times 3} & \\
0 & & &
\end{pmatrix}
\end{eqnarray}
where ${\cal O}_{3\times3}$ is a $3\times 3$ orthogonal matrix which we can construct using the eigenvectors (normalized to be a unit vector) $\vec{x}_{i}$ of the mass matrix $\mu^{2}$,
\begin{eqnarray}
{\cal O}_{3\times 3} = \left(
\begin{array}{r|r|r} 
& & \\
\vec{x}_{1} & \vec{x}_{2} & \vec{x}_{3}\\
& &
\end{array} \right).
\end{eqnarray}
Analytically, we also can decompose the orthogonal matrix as given by in the Refs. \cite{Kronenburg:2004,Kronenburg:2015},
\begin{eqnarray} \label{decomposition}
\mathcal{O}_{3\times 3}  = R_{1}(\theta_{1}) \cdot R_{2}(\theta_{2}) \cdot R_{3}(\theta_{3})
\end{eqnarray}
where
\begin{eqnarray}
& & 
R_{1}(\theta_{1}) = 
\begin{pmatrix}
1 & 0 & 0 \\
0 & \cos\theta_{1} & -\sin\theta_{1} \\
0 & \sin\theta_{1} & \cos\theta_{1} \\
\end{pmatrix}, \ R_{2}(\theta_{2}) = 
\begin{pmatrix}
\cos\theta_{2} & 0 & \sin\theta_{2} \\
0 & 1 & 0 \\
-\sin\theta_{2} & 0 & \cos\theta_{2} \\
\end{pmatrix}, \
R_{3}(\theta_{3}) = 
\begin{pmatrix}
\cos\theta_{3} & -\sin\theta_{3} & 0 \\
\sin\theta_{3} & \cos\theta_{3} & 0 \\
0 & 0 & 1 \\
\end{pmatrix}. \nonumber \\
\end{eqnarray}
The method of calculating $ \theta_{i}~(i=1,2,3) $ in \cite{Kronenburg:2004, Kronenburg:2015} is also reviewed in Section~\ref{app.angles}.

Finally, the gauge eigenstates $(\hat{A}, \hat{Z}, \hat{X}_1,\hat{X}_2)$ are related with the mass eigenstates $(A, Z_1, Z_2, Z_3)$ as
\begin{eqnarray}
\begin{pmatrix} 
\hat{A} \\
\hat{Z} \\
\hat{X}_1 \\
\hat{X}_2
\end{pmatrix} & = & \begin{pmatrix}
1 & 0 & -\hat{c}_W t_\alpha & \hat{c}_W k/D \\
0 & 1 & \hat{s}_W t_\alpha &  -\hat{s}_W k/D\\
0& 0 & 1/c_\alpha & q/D \\
0 & 0& 0& c_\alpha/D
\end{pmatrix}\begin{pmatrix}
1 & 0 & 0 & 0 \\
0 & & & \\
0 & & {\cal O}_{3\times 3}^T & \\
0 & & &
\end{pmatrix} \begin{pmatrix} 
A\\
Z_1 \\
Z_2\\
Z_3
\end{pmatrix}
\end{eqnarray}
or inverted relation is given as
\begin{eqnarray}
\begin{pmatrix} 
A \\
Z_1 \\
Z_2 \\
Z_3
\end{pmatrix} & = & \begin{pmatrix}
1 & 0 & 0 & 0 \\
0 & & & \\
0 & & {\cal O}_{3\times 3} & \\
0 & & &
\end{pmatrix}
\begin{pmatrix}
1 & 0 & \hat{c}_W s_\alpha & \hat{c}_W s_\beta \\
0 & 1 & -\hat{s}_W s_\alpha &  -\hat{s}_W s_\beta \\
0& 0 & c_\alpha & (s_\gamma -s_\alpha s_\beta)/c_\alpha \\
0 & 0& 0& D/c_\alpha
\end{pmatrix}
\begin{pmatrix} 
\hat{A} \\
\hat{Z} \\
\hat{X}_1 \\
\hat{X}_2
\end{pmatrix}.
\end{eqnarray}

The approximated form of mass matrix $\mu^2$ and corresponding $\mathcal{O}_{3 \times 3} $ in the limit $ m_{i}^{2} \ll \hat{M}_{Z}^{2}, \hat{M}_{X_{j}}^{2}$ and $ \alpha,\beta,\gamma \ll 1 $ is given in the Ref. \cite{Heeck:2011md} as
\begin{eqnarray}
\mu^{2} & \simeq & \begin{pmatrix}
\hat{M}_Z^2 &\hat{M}_Z^2 \hat{s}_W \alpha +m_{1}^2 & \hat{M}_Z^2 \hat{s}_W \beta +m_{2}^2 \\ 
\hat{M}_Z^2 \hat{s}_W \alpha +m_{1}^2 & \hat{M}_{X_1}^2 &-\hat{M}_{X1}^2\gamma +m_{3}^2\\ 
\hat{M}_Z^2 \hat{s}_W \beta +m_{2}^2 & -\hat{M}_{X1}^2\gamma +m_{3}^2 & \hat{M}_{X_2}^2
\end{pmatrix}, \\
\mathcal{O}_{3 \times 3} & \simeq &
\begin{pmatrix}
1& \frac{\hat{s}_{W}\alpha \hat{M}_{X1}^{2}+m_{1}^{2}}{\hat{M}_{X1}^{2}-\hat{M}_{Z}^{2}}&\frac{\hat{s}_{W}\beta \hat{M}_{X2}^{2}+m_{2}^{2}}{\hat{M}_{X2}^{2}-\hat{M}_{Z}^{2}} \\
-\frac{\hat{s}_{W}\alpha \hat{M}_{X1}^{2}+m_{1}^{2}}{\hat{M}_{X1}^{2}-\hat{M}_{Z}^{2}} &1 & -\frac{\gamma\hat{M}_{X2}^{2}-m_{3}^{2}}{\hat{M}_{X2}^{2}-\hat{M}_{1}^{2}}\\
-\frac{\hat{s}_{W}\beta \hat{M}_{X2}^{2}+m_{2}^{2}}{\hat{M}_{X2}^{2}-\hat{M}_{Z}^{2}} &\frac{\gamma\hat{M}_{X2}^{2}-m_{3}^{2}}{\hat{M}_{X2}^{2}-\hat{M}_{1}^{2}} &1
\end{pmatrix},
\end{eqnarray}
respectively. The diagonalization leads to the following connection between gauge and mass eigenstates
\begin{eqnarray}
\begin{pmatrix} 
\hat{A} \\
\hat{Z} \\
\hat{X}_1 \\
\hat{X}_2
\end{pmatrix} & \simeq & \begin{pmatrix}
1 & 0 & -\hat{c}_{W}\alpha & -\hat{c}_{W}\beta \\
0 & 1& \frac{\hat{s}_{W}\alpha \hat{M}_{X1}^{2}+m_{1}^{2}}{\hat{M}_{X1}^{2}-\hat{M}_{Z}^{2}}&\frac{\hat{s}_{W}\beta \hat{M}_{X2}^{2}+m_{2}^{2}}{\hat{M}_{X2}^{2}-\hat{M}_{Z}^{2}} \\
0 &-\frac{\hat{s}_{W}\alpha \hat{M}_{X1}^{2}+m_{1}^{2}}{\hat{M}_{X1}^{2}-\hat{M}_{Z}^{2}} &1 & -\frac{\gamma\hat{M}_{X2}^{2}-m_{3}^{2}}{\hat{M}_{X2}^{2}-\hat{M}_{1}^{2}}\\
0 &-\frac{\hat{s}_{W}\beta \hat{M}_{X2}^{2}+m_{2}^{2}}{\hat{M}_{X2}^{2}-\hat{M}_{Z}^{2}} &\frac{\gamma\hat{M}_{X2}^{2}-m_{3}^{2}}{\hat{M}_{X2}^{2}-\hat{M}_{1}^{2}} &1
\end{pmatrix} \begin{pmatrix}
A \\
Z_1 \\
Z_2 \\
Z_3
\end{pmatrix}.
\end{eqnarray}
The analytic method of calculating $\mathcal{O}_{3\times 3}$ without assumptions is presented in the next section.

\subsection{Formulae for eigenvalues and angle parameters~($\theta_{1},\theta_{2},\theta_{3}$)}\label{app.angles}
In this section, we summarize the formulae given in the Refs. \cite{Kronenburg:2004,Kronenburg:2015} of calculating the eigenvalues and the orthogonal matrix $ \mathcal{O}_{3 \times 3} $.

Let us define $p$, $q$, and $\Delta$ as
\begin{eqnarray}
p & = & \frac{1}{2} \left[ (\mu_{11}^{2}-\mu_{22}^{2})^{2} + (\mu_{11}^{2}-\mu_{33}^{2})^{2}+(\mu_{22}^{2}-\mu_{33}^{2})^{2}	\right] + 3[(\mu_{12}^{2})^{2}+(\mu_{13}^{2})^{2}+(\mu_{23}^{2})^{2}], \ \ \ \ \ \ \ \\
q & = &18(\mu_{11}^{2}\mu_{22}^{2}\mu_{33}^{2}
+3\mu_{12}^{2}\mu_{13}^{2}\mu_{23}^{2}
+2[(\mu_{11}^{2})^{3}+(\mu_{22}^{2})^{3}+(\mu_{33}^{2})^{3}] \nonumber \\ 
& & +9(\mu_{11}^{2}+\mu_{22}^{2}+\mu_{33}^{2})[(\mu_{12}^{2})^{2}+(\mu_{13}^{2})^{2}+(\mu_{23}^{2})^{2}] \nonumber\\
& & -3(\mu_{11}^{2}+\mu_{22}^{2})(\mu_{11}^{2}+\mu_{33}^{2})(\mu_{22}^{2}+\mu_{33}^{2})
-27[\mu_{11}^{2}(\mu_{23}^{2})^{2}+\mu_{22}^{2}(\mu_{13}^{2})^{2}+\mu_{33}^{2}(\mu_{12}^{2})^{2}], \nonumber \\
\end{eqnarray}
and
\begin{eqnarray}
\Delta & = & \arccos \left(\frac{q}{2\sqrt{p^{3}}}\right)
\end{eqnarray}
respectively. $\Delta$ is chosen to be a value of the range $\left( -\pi/2, \pi/2 \right)$.

The mass spectrum (i.e. eigenvalues) $ (\mu_{1}^2, \mu_{2}^2,\mu_{3}^2) $ of the 3 by 3 symmetric matrix $ \mu^{2} $ can be represented as following:
\begin{eqnarray}
\mu_{1}^2 & = & \frac{1}{3} \left[(\mu_{11}^{2}+\mu_{22}^{2}+\mu_{33}^{2}) + 2\sqrt{p} \cos \left( \frac{\Delta}{3}\right) \right] \\
\mu_{2}^2 & = & \frac{1}{3} \left[(\mu_{11}^{2}+\mu_{22}^{2}+\mu_{33}^{2}) + 2\sqrt{p} \cos \left( \frac{\Delta+2\pi}{3}\right) \right] \\
\mu_{3}^2 & = & \frac{1}{3} \left[(\mu_{11}^{2}+\mu_{22}^{2}+\mu_{33}^{2}) + 2\sqrt{p} \cos \left( \frac{\Delta-2\pi}{3}\right) \right].
\end{eqnarray}
In our convention, $\mu_{1}^{2} \geq \mu_{3}^{2} \geq \mu_{2}^{2}$ when $\Delta \geq 0$, and $\mu_{1}^{2} \geq \mu_{2}^{2} \geq \mu_{3}^{2}$ when $\Delta < 0$.

The angles $\theta_{2}$ and $\theta_{3}$ introduced in Eq.~(\ref{decomposition}) are represented by
\begin{eqnarray}
& & \cos \theta_2 = \Theta_2, \ \ \  
\cos \theta_3=\Theta_3,
\end{eqnarray}
where
\begin{eqnarray}
\Theta_2 & = & \sqrt{\frac{(\mu^{2}_{12})^{2}+(\mu^{2}_{13})^{2}+(\mu^{2}_{11}-\mu_{3}^2)(\mu^{2}_{11}+\mu_{3}^2-\mu_{1}^2-\mu_{2}^2)}{(\mu_{1}^2-\mu_{3}^2)(\mu_{3}^2-\mu_{1}^2)}}, \\
\Theta_3 & = & \sqrt{\frac{\mu^{2}_{11}-\mu_{3}^2+(\mu_{3}^2-\mu_{2}^2)\Theta_2^2}{(\mu_{1}^2-\mu_{2}^2)\Theta_2^2}}.
\end{eqnarray}
The sign of two angles should be determined after $\theta_{1}$ is specified. For $\theta_{1}$, one need to divide the cases. Let the following 2-dimensional auxilliary vectors to be defined:
\begin{eqnarray}
f_{1} & = & \begin{pmatrix}
\mu^{2}_{12}\\-\mu^{2}_{13}
\end{pmatrix},~~~~~
f_{2}=\begin{pmatrix}
\mu^{2}_{22}-\mu^{2}_{33}\\-2\mu^{2}_{23}
\end{pmatrix}, \nonumber \\
g_{1} & = & \begin{pmatrix}
\frac{1}{2}(\mu_{1}^2-\mu_{2}^2)\cos\theta_{2} \sin 2\theta_{3} \\\frac{1}{2}\left[(\mu_{1}^2-\mu_{2}^2)\Theta_3^2+\mu_{2}^2-\mu_{3}^2\right]\sin 2\theta_{2}
\end{pmatrix}, \nonumber \\
g_{2} & = & \begin{pmatrix}
(\mu_{1}^2-\mu_{2}^2)\left[1+(\Theta_2^2-2)\Theta_3^2\right]+(\mu_{2}^2-\mu_{3}^2)\Theta_2^2   \\(({\mu_{1}^{2}})^2-{(\mu_{2}^{2}})^2)\sin\theta_{2} \sin 2 \theta_{3}
\end{pmatrix}
\end{eqnarray}
with a properties of
\begin{eqnarray}
& & \vert g_{1} \vert = \vert f_{1} \vert,~~~\vert g_{2} \vert = \vert f_{2} \vert
\end{eqnarray}
and the $\theta_{1}$ satisfies
\begin{eqnarray}
g_{1} & = & R(\theta_{1}) f_{1} \\
g_{2} & = & R(2\theta_{1}) f_{2}
\end{eqnarray}
where $R(\phi)=\begin{pmatrix}
\cos\phi & \sin\phi \\
-\sin\phi & \cos\phi
\end{pmatrix}$. This means that $\theta_{1}$ is equal to the angle between $g_{1}$ and $f_{1}$ and half the angle between $g_{2}$ and $f_{2}$. Let angle($ v $) be the angle of a 2-dimensional vector $v$ with respect to the vector $(1,0)^{\rm T}$, which is equivalently the angle with positive $ x $ axis. We also define
\begin{eqnarray}
\phi_{i} & = & {\rm angle}(f_{i})~~~(i=1,2)
\end{eqnarray}
and $\theta_{1}$ is given by
\begin{eqnarray}
\theta_{1} & = &
\begin{cases} 
\theta_{1(1)}={\rm angle}(R(\phi_{1})\cdot g_{1})~~~~~{\rm if}~~~\vert f_{1} \vert \geq \vert f_{2} \vert  \\
\theta_{1(2)}=\frac{1}{2}{\rm angle}(R(\phi_{2})\cdot g_{2})~~~{\rm if}~~~\vert f_{1} \vert < \vert f_{2} \vert\\
\end{cases}.
\end{eqnarray}
The sign combination of $\theta_{2}$ and $\theta_{3}$ is determined to be with the smallest difference between $\theta_{1(1)}$ and $\theta_{1(2)}$.

\section{One-loop FCNC induced by a light gauge boson coupled to third generation quarks}
\label{appendix_oneloop_FCNC}

In model II (Section \ref{subsec:Model2_LmuLtau_B3Ltau_w_RHnu}), $X$ gauge boson coupled to a combination of fermion numbers $(L_\mu - L_\tau) + \epsilon (B_3 - L_\tau)$ generally induces the monopole and the dipole FCNC terms at one-loop level as
\begin{eqnarray}
\Delta \mathcal{L}_{dsX} & \sim & \frac{\epsilon g_X}{3} \frac{g^2 (V_{ts}^* V_{td})}{16\pi^2} \Bigl \{ F_1(x_t) (\bar{d} \gamma^\mu P_L s) X_\mu + \frac{1}{m_W^2} F_2 (x_t) (q^2 g^{\mu \nu} - q^\mu q^\nu) (\bar{d} \gamma_\mu P_L s) X_\nu \nonumber \\
& & + \frac{m_s}{m_W^2} F_3 (x_t) (\bar{d} \sigma^{\mu \nu} P_L s) q_\nu X_\mu + \frac{m_d}{m_W^2} F_4 (x_t) (\bar{d} \sigma^{\mu \nu} P_R s) q_\nu X_\mu \Bigr \}
\end{eqnarray}
which $F_i (x_t)$ $(i=1,2,3,4)$ are the (dimensionless) function of order 1. At low energy, $q$ is the momentum of the produced $X$ gauge boson and $q^2 = m_X^2 \ll m_t^2 , m_W^2$. We focus on the $F_1(x_t) (\bar{d} \gamma^\mu P_L s) X_\mu$ term, which is dominant in our case.

In Fig. \ref{fig_FCNC_B3_dsX_diagrams}, we show the diagrams which contribute to FCNC vertex and the amplitudes are given by $i \mathcal{M}_{a,b}^{B_3} = \epsilon_\mu^{X*}(q) \cdot \bar{d} (i \Gamma_{a,b}^{(B_3)\mu}) s $. For the diagram $a$ (Fig. \ref{fig_FCNC_B3_dsX_diagram_a}), one obtains one-loop amplitude as
\begin{eqnarray}
\bar{d} (i \Gamma_a^{(B_3)\mu}) s & = & \int \frac{d^d k}{(2\pi)^d} \bar{d} \left [ \left ( \frac{ig}{\sqrt{2}} \gamma^\rho P_L V_{ts}^* \right ) \frac{i (\slashed{p}_2 - \slashed{k} + m_t)}{(p_2 - k)^2 - m_t^2} \left ( -i \frac{1}{3} \epsilon g_X \gamma^\mu \right ) \right . \nonumber \\ 
& & \left. \times \frac{i (\slashed{p}_1 - \slashed{k} + m_t)}{(p_1 - k)^2 - m_t^2} \left (\frac{ig}{\sqrt{2}} \gamma^\nu P_L V_{td} \right ) \right ] s \times \frac{- i g_{\rho \nu}}{k^2 - m_W^2}
\end{eqnarray}
and the vertex correction $\Gamma_a^{(B_3)\mu}$ is explicitly given by \cite{Aoki:1982ed}
\begin{eqnarray}
i\Gamma_a^{(B_3)\mu} & = & \left (- \frac{g^2}{2} V_{t2}^* V_{t1} \cdot \frac{i}{4} \cdot \frac{\epsilon g_X}{3} \right ) \left \{ \frac{m_t^2}{2\pi^2} \gamma^\mu P_L \int_0^1 \int_0^1 dx dy \ \frac{y}{\Delta} \right. \nonumber \\
& & + \frac{1}{2\pi^2} P_R \int_0^1 \int_0^1 \frac{dx dy}{\Delta} y (\slashed{p}_1 - y \bar{\slashed{p}}) \gamma^\mu (\slashed{p}_2 - y \bar{\slashed{p}} ) \nonumber \\
& & \left. + \frac{1}{4\pi^2} \gamma^\mu P_L \Bigl ( C_\epsilon - 2 \int_0^1 \int_0^1 dx dy \ y \ln \Delta \Bigr ) \right \}
\end{eqnarray}
where $\epsilon = 4 - d$ for the dimensional regularization and
\begin{eqnarray}
C_\epsilon & = & \frac{1}{\epsilon} - \gamma_E + \ln 4\pi - 2, \\
\Delta & = & (1-y) m_W^2 + y \{ m_t^2 - q^2 x(1-x) \} - y (1-y) \bar{p}^2, \\
\bar{p} & = & (1-x)p_1 + xp_2.
\end{eqnarray}

In the limit $m_d^2,  m_s^2, q^2(=m_X^2) \ll m_t^2, m_W^2$, we approximate the loop-induced vertex as
\begin{eqnarray}
\Gamma_a^{(B_3)\mu} & \simeq & \left (- \frac{\epsilon g_X}{3} \right) \frac{g^2 (V_{t2}^* V_{t1})}{16\pi^2} \left (A(x_t) + B(x_t) \right ) \gamma^\mu P_L
\end{eqnarray}
where
\begin{eqnarray}
A(x_t) & \equiv & m_t^2 \int_0^1 dy \frac{y}{y(m_t^2 - m_W^2) + m_W^2} = \frac{x_t (x_t-1) - x_t \ln x_t}{(x_t-1)^2}, \\
B(x_t) & \equiv & - \int_0^1 dy \ y \cdot \ln [y(m_t^2 - m_W^2) + m_W^2] = \frac{x_t (x_t - 1) - x_t^2 \ln x_t}{2(x_t - 1)^2},
\end{eqnarray}
and $x_t \equiv m_t^2 /m_W^2$. Similarly, in the same limit, the effective vertex $\Gamma_b^{(B_3)\mu}$ for the diagram $b$ (Fig. \ref{fig_FCNC_B3_dsX_diagram_b}) also can be approximately written as
\begin{eqnarray}
\Gamma_b^{(B_3)\mu} & \simeq & \left (- \frac{\epsilon g_X}{3} \right ) \frac{g^2 (V_{ts}^* V_{td})}{16\pi^2} \left \{ \frac{x_t}{2} A(x_t) + \frac{x_t}{2} B(x_t) - \frac{x_t}{4} \right \} \gamma^\mu P_L.
\end{eqnarray}

As a result, we get the loop-induced FCNC vertex as
\begin{eqnarray}
\Gamma^{(B_3)\mu} \equiv \Gamma_a^{(B_3)\mu} + \Gamma_b^{(B_3)\mu} & \simeq & \left (\frac{\epsilon g_X}{3} \right ) \frac{g^2 (V_{ts}^* V_{td})}{16\pi^2} F_1 (x_t) \gamma^\mu P_L
\end{eqnarray}
where
\begin{eqnarray}
F_1(x_t) & = & - A(x_t) - B(x_t) - \frac{x_t}{2} A(x_t) - \frac{x_t}{2} B(x_t) + \frac{x_t}{4} \nonumber \\
& = & \frac{7x_t - 5 x_t^2 - 2 x_t^3 + x_t(x_t + 2)^2 \ln x_t}{4(x_t - 1)^2}
\end{eqnarray}
is the loop function of order 1.



\bibliographystyle{JHEP}
\bibliography{biblio}

\end{document}